\documentclass{aa}
\usepackage{graphicx,txfonts}

\begin{document}

\title{
        $\alpha$-, $r$-, and $s$-process element trends in the \\
        Galactic thin and thick disks\thanks{Based on observations
        collected at the Nordic Optical Telescope on La Palma, Spain,
        and at the European Southern Observatories on La Silla and
        Paranal, Chile, Proposals \#\,65.L-0019(B), 67.B-0108(B),
        69.B-0277.}\fnmsep\thanks{Tables~\ref{tab:atomdata},
        \ref{tab:ages}, and \ref{tab:abundances} are only available in
        electronic form at the CDS via anonymous ftp to
        \texttt{cdsarc.u-strasbg.fr} (130.79.128.5) or via
        \texttt{http://cdsweb.u-strasbg.fr/cgi-bin/qcat?J/A+A/}} 
    }

\author{T. Bensby\inst{1}\fnmsep\inst{2}
 \and S. Feltzing\inst{1}
 \and I. Lundstr\"om\inst{1}
 \and I. Ilyin\inst{3}}

\offprints{Thomas Bensby}

\institute{
        Lund Observatory, Box 43, SE-221\,00 Lund, Sweden \\
        \email{sofia@astro.lu.se; ingemar@astro.lu.se}
        \and
        Department of Astronomy, 921 Dennison Building, University of Michigan,
        Ann Arbor, MI~48109-1090, USA\\ 
        \email{tbensby@umich.edu}
        \and
        Astrophysical Institute Potsdam,
        An der Sternwarte 16,
        D-14482 Potsdam, Germany \\
        \email{ilyin@aip.de}
        }

\date{Received 25 February 2004 / Accepted 2 December 2004}

%==============================================================================
\abstract{
        From a detailed elemental abundance analysis of 102 F and G
        dwarf stars we present abundance trends in the Galactic thin
        and thick disks for 14 elements (O, Na, Mg, Al, Si, Ca, Ti,
        Cr, Fe, Ni, Zn, Y, Ba, and Eu).  Stellar parameters and
        elemental abundances (except for Y, Ba and Eu) for 66 of the
        102 stars were presented in our previous studies (Bensby et
        al.~\cite{bensby},~\cite{bensby_syre}). The 36 stars that are
        new in this study extend and confirm our previous results and
        allow us to draw further conclusions regarding abundance
        trends. The s-process elements Y and Ba, and the r-element Eu
        have also been considered here for the whole sample for the
        first time.  With this new larger sample we now have the
        following results:
        1) Smooth and distinct abundance trends that for the thin and
           thick disks are clearly separated;
        2) The $\alpha$-element trends for the thick disk show typical
           signatures from the enrichment of SN\,Ia;
        3) The thick disk stellar sample is in the mean older than the
           thin disk stellar sample;
        4) The thick disk abundance trends are invariant with
           galactocentric radii ($R_{\rm m}$);
        5) The thick disk abundance trends appear to be invariant with
           vertical distance ($Z_{\rm max}$) from the Galactic plane.
        Adding further evidence from the literaure we argue that a
        merger/interacting scenario with a companion galaxy to produce
        a kinematical heating of the stars (that make up today's thick
        disk) in a pre-existing old thin disk is the most likely
        formation scenario for the Galactic thick disk. \\ The 102
        stars have $\rm -1 \lesssim [Fe/H] \lesssim +0.4$ and are all
        in the solar neighbourhood. Based on their kinematics they
        have been divided into a thin disk sample and a thick disk
        sample consisting of 60 and 38 stars, respectively. The
        remaining 4 stars have kinematics that make them kinematically
        intermediate to the two disks.  Their chemical abundances also
        place them in between the two disks.  Which of the two disk
        populations these 4 stars belong to, or if they form a
        distinct population of their own, can at the moment not be
        settled.  The 66 stars from our previous studies were observed
        with the FEROS spectrograph on the ESO 1.5-m telescope and the
        CES spectrograph on the ESO 3.6-m telescope. Of the 36 new
        stars presented here 30 were observed with the SOFIN
        spectrograph on the Nordic Optical Telescope on La Palma, 3
        with the UVES spectrograph on VLT/UT2, and 3 with the FEROS
        spectrograph on the ESO 1.5-m telescope. All spectra have high
        signal-to-noise ratios (typically $S/N\gtrsim 250$) and high
        resolution ($R\sim 80\,000$, 45\,000, and 110\,000 for the
        SOFIN, FEROS, and UVES spectra, respectively).
 }

\maketitle

%==============================================================================
\section{Introduction}

During the last few years several studies have used detailed abundance
analysis in order to establish the chemical properties of the thick
disk stellar population (e.g., Bensby et al.~\cite{bensby},
\cite{bensby_syre}; Feltzing et al.~\cite{bensby_letter}; Reddy et
al~\cite{reddy}; Tautvai\v sien\.e et al.~\cite{tautvaisiene};
Mashonkina \& Gehren~\cite{mashonkina}; Gratton et
al.~\cite{gratton2}; Prochaska et al.~\cite{prochaska}; Chen et
al.~\cite{chen}; Fuhrmann~\cite{fuhrmann}). 
Although the various studies take different approaches to defining
the stellar  samples and though some of them are only concerned with one 
of the disks, there is a general agreement on the following: 1) The thick 
disk is, at a given [Fe/H], more enhanced in the $\alpha$-elements than the 
thin disk; 2) The abundance trend in the thin disk is a gentle slope, and 
3) The solar neighbourhood thick disk stars that have been studied so far 
are all old.

The aim of the present study is to verify and extend these results
and to add  new
elements into the discussion; the $r$-process element europium (Eu) and
the two $s$-process elements yttrium (Y) and barium (Ba). 
By studying Eu and Ba  Mashonkina \& Gehren~(\cite{mashonkina})
found that AGB stars have contributed to the chemical enrichment
of the thick disk. By including these elements we
will be able to confirm this important finding.  These elements will 
also be combined with the $\alpha$-elements, in particular our oxygen
abundances from Bensby et al.~(\cite{bensby_syre}), to shed new light on 
the chemical enrichment. 

The paper is organized as follows. In Sect.~\ref{sec:sample} we
describe the stellar sample.  Section~\ref{sec:sofin} describes the
observations and the data reductions.  Sections~\ref{sec:atmospheres}
and \ref{sec:analysis} briefly describe the stellar model atmospheres
and the elemental abundance determination. The interested
reader is refered to Bensby et al.~(\cite{bensby}) for a detailed
discussion. Section~\ref{sec:ages}
describes how we determined stellar ages.  The resulting elemental
abundance trends are then presented in Sect.~\ref{sec:results} and
combined with the results from Bensby et al.~(\cite{bensby},
\cite{bensby_syre}) for an extended discussion.  Conclusions and a
final summary are given in in Sect.~\ref{sec:summary}.  The paper ends
with an Appendix that includes a discussion of the assumptions made about
the parameters that are used in the kinematical selection criteria for
the stellar samples.

%------------------------------------------------------------------------------
\begin{figure}
\resizebox{\hsize}{!}{\includegraphics{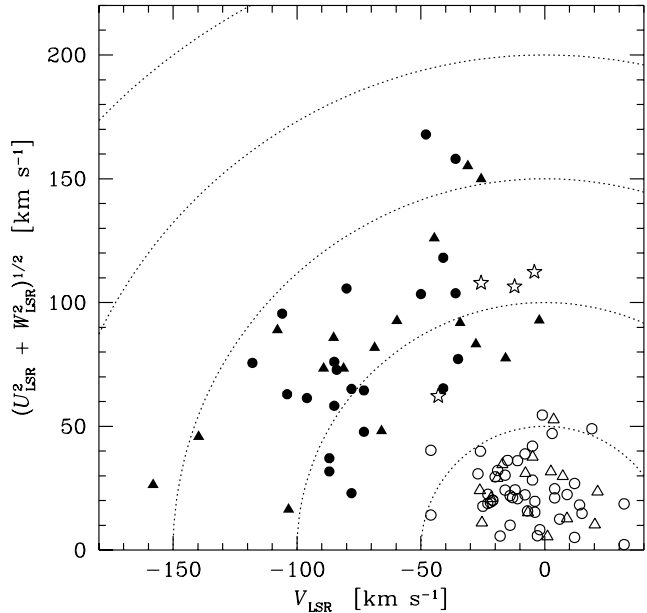}}
\caption{
        Toomre diagram for the full stellar sample (102 stars). 
        Thick and thin disk stars
        are marked by filled and open symbols, respectively.
        Stars that have been observed with SOFIN or UVES are marked by
        triangles and those from Bensby et al.~(\cite{bensby}) 
        are marked by circles.
        ``Transition objects" are marked by ``open stars".
         }
\label{fig:toomre}
\end{figure}
%------------------------------------------------------------------------------

%------------------------------------------------------------------------------
\begin{table}
\centering
\caption{
        Kinematical data for the new stellar sample. Column~1 gives the
        Hipparcos number; col.~2 gives the radial velocities 
        (as measured by us); 
        cols.~3--5 give the space velocities relative to the local standard
        of rest; cols.~6 and 7 give the calculated $TD/D$ and $TD/H$ ratios.
        The radial velocity for HIP~116740 have been taken from 
        Barbier-Brossat et al.~(\cite{barbier}).
        }
\centering %\scriptsize%\tiny
\setlength{\tabcolsep}{1.4mm}
\begin{tabular}{rrrrrrr}
\hline \hline\noalign{\smallskip}
           HIP                               
        &  $v_{\rm r}$                       
        &  \multicolumn{1}{c}{$U_{\rm LSR}$} 
        &  \multicolumn{1}{c}{$V_{\rm LSR}$} 
        &  \multicolumn{1}{c}{$W_{\rm LSR}$} 
        &  $TD/D$                                    
        &  $TD/H$                              \\
\noalign{\smallskip}
        &  \multicolumn{4}{c}{~~~----------~~ [km~s$^{-1}$] ~~----------} 
        & 
        & \\
\noalign{\smallskip}                                
\hline                                                                     
\noalign{\smallskip}                                              
\multicolumn{7}{c}{\bf THIN DISK STARS}                                 \\   
    699 & $ -14.9$ &  $  37.6$ & $  -4.9$ & $  -0.8$ &  0.013 &  $>$999 \\   
    910 & $  15.0$ &  $  29.0$ & $  -7.8$ & $ -11.1$ &  0.014 &  $>$999 \\   
   2235 & $  -9.9$ &  $  23.2$ & $ -26.2$ & $  -6.7$ &  0.018 &  $>$999 \\   
   2787 & $  11.7$ &  $ -16.5$ & $ -21.7$ & $ -12.9$ &  0.017 &  $>$999 \\   
   3909 & $   7.6$ &  $  31.4$ & $   2.5$ & $  -4.8$ &  0.013 &  $>$999 \\   
   5862 & $  14.3$ &  $ -34.1$ & $ -17.1$ & $  -6.0$ &  0.016 &  $>$999 \\   
  10306 & $   5.9$ &  $ -10.3$ & $  -6.8$ & $  11.2$ &  0.011 &  $>$999 \\   
  15131 & $   6.7$ &  $   3.6$ & $  20.2$ & $  -9.6$ &  0.017 &  $>$999 \\   
  18833 & $ -29.5$ &  $  50.8$ & $   3.6$ & $ -14.0$ &  0.027 &  $>$999 \\   
  88945 & $ -14.4$ &  $  -4.3$ & $   1.1$ & $  -3.3$ &  0.009 &  $>$999 \\  
  92270 & $   3.6$ &  $  12.0$ & $   9.0$ & $   4.4$ &  0.011 &  $>$999 \\ 
  93185 & $ -40.4$ &  $ -29.0$ & $ -19.1$ & $   2.2$ &  0.014 &  $>$999 \\ 
  96258 & $   1.0$ &  $  29.3$ & $   7.3$ & $  -5.4$ &  0.013 &  $>$999 \\ 
 107975 & $  19.3$ &  $  23.6$ & $  21.3$ & $  -0.3$ &  0.018 &  $>$999 \\ 
 113174 & $ -28.1$ &  $  -1.8$ & $ -25.4$ & $  10.9$ &  0.017 &  $>$999 \\ 
\noalign{\smallskip}
\multicolumn{7}{c}{\bf THICK DISK STARS}                                \\ 
  11309 & $  -8.5$ &  $  73.3$ & $ -34.2$ & $ -55.3$ &  11.7  &  $>$999 \\ 
  12306 & $ -99.7$ &  $ 146.0$ & $ -25.7$ & $ -33.9$ &  46.1  &  819    \\ 
  15510 & $  90.3$ &  $ -69.4$ & $ -89.3$ & $ -24.0$ &  73.7  &  765    \\ 
  16788 & $ -29.1$ &  $  78.0$ & $  -2.2$ & $ -50.3$ &  2.7   &  $>$999 \\ 
  18235 & $ 100.7$ &  $ -17.0$ & $-158.1$ & $ -20.1$ & $>$999 &  13.4   \\ 
  20242 & $ -31.6$ &  $  90.5$ & $ -59.7$ & $ -19.8$ &  3.3   &  $>$999 \\ 
  21832 & $ 111.5$ &  $-124.6$ & $ -44.6$ & $  18.9$ &  7.3   &  $>$999 \\ 
  26828 & $  75.8$ &  $ -61.4$ & $ -81.2$ & $  40.2$ &  80.9  &  838    \\ 
  36874 & $-134.9$ &  $ 149.7$ & $ -31.1$ & $ -40.7$ &  179   &  584    \\ 
  37789 & $  -5.4$ &  $   4.5$ & $-103.4$ & $ -15.8$ &  115   &  657    \\ 
  40613 & $ 113.4$ &  $ -28.4$ & $-139.7$ & $ -35.9$ & $>$999 &  41.0   \\ 
  44075 & $ 119.5$ &  $ -38.2$ & $ -85.2$ & $  76.8$ & $>$999 &  198    \\ 
  44860 & $  66.8$ &  $ -58.9$ & $ -68.7$ & $ -56.7$ &  173   &  796    \\ 
 112151 & $  -8.3$ &  $ -10.9$ & $ -65.9$ & $ -46.9$ &  9.6   &  $>$999 \\ 
 116421 & $-112.1$ &  $ -55.1$ & $-107.9$ & $  69.8$ & $>$999 &  78.2   \\ 
 116740 & $ -32.0$ &  $ -68.5$ & $ -27.8$ & $  47.3$ &  1.94  &  $>$999 \\ 
 118010 & $   3.3$ &  $ -45.9$ & $ -15.8$ & $  62.5$ &   8.0  &  $>$999 \\ 
\noalign{\smallskip}
\multicolumn{7}{c}{\bf ``TRANSITION OBJECTS"}                           \\ 
   3170 & $   1.9$ &  $-102.8$ & $ -12.2$ & $ -27.6$ &  0.71  &  $>$999 \\ 
  44441 & $  72.8$ &  $-110.5$ & $  -4.2$ & $ -20.2$ &  0.62  &  $>$999 \\ 
  95447 & $-101.3$ &  $-107.1$ & $ -25.7$ & $ -13.3$ &  0.55  &  $>$999 \\ 
 100412 & $  17.0$ &  $  46.1$ & $ -43.0$ & $  41.8$ &  1.04  &  $>$999 \\ 
\noalign{\smallskip}
\hline
\end{tabular}
\label{tab:kinematics}
\end{table}
%------------------------------------------------------------------------------

%==============================================================================
\section{Stellar sample} \label{sec:sample}

Given that we, in principle, never can select a sample of local thick disk
stars that is guaranteed to be completely free from intervening thin disk 
stars, we argue that we should keep the selection criteria as simple and as
transparent as possible. In this sense the simplest and most honest 
selection is based only on the kinematics of the stars. 
This is also the least model dependent method.
The selection method we used is described and 
discussed in Bensby et al.~(\cite{bensby}), 
see also Bensby et al.~(\cite{bensby_syre}, \cite{bensby_amr}).

Our study contains two major stellar samples. They have been defined
to kinematically resemble the thin and thick Galactic disks, respectively. 
As mentioned the criteria and method are described in Bensby 
et al.~(\cite{bensby}).  However, while we
then used a 6\,\% for the normalization of the thick disk stars in the
solar neighbourhood we here use 10\,\% (and consequently the
normalization for the thin disk population is 90\,\%). Reasons for
this are given in the Appendix.

Our total stellar sample contains 38 thick disk stars and 60 thin disk
stars. The new sample contains 17
thick disk stars, 15 thin disk stars (all with [Fe/H]\,$< 0$) and a
further 4 stars with kinematics intermediate between the thin and
thick disks. By intermediate we mean that they can be classified
either as thin disk or as thick disk stars depending on the value for
the solar neighbourhood thick disk stellar density.  Using a value of
6\,\% will classify them as thin disk stars whereas a value of 14\,\%
will classify them as thick disk stars. Due to this ambiguity we will
treat these stars (HIP~3170, HIP~44441, HIP~95447, and HIP~100412)
separately from the two other samples and label them as ``transition
objects". The remaining 21 thick disk and 45 thin disk stars were analyzed
in Bensby et al.~(\cite{bensby}).

Radial velocities were determined for 35 of the 36 stars in the new
sample.  Good agreement to the radial velocities in the compilation by
Barbier-Brossat et al.~(\cite{barbier}) (which were used when selecting
the stars for the observations) is generally found, with the exception
of HIP 18833 where our radial velocity is 13\,km\,s$^{-1}$ lower. The
average difference for the other stars is $+0.4\pm 3$\,km\,s$^{-1}$,
with our measurements giving the larger values.  For one star (HIP
116740) we adopted the radial velocity as given in Barbier-Brossat et
al.~(\cite{barbier}) since there was an offset in the wavelength shift
between the red and the blue settings in the SOFIN spectra for this star 
(compare Table~\ref{tab:sofin}).  All kinematical properties and the 
calculated $TD/D$ and $TD/H$ ratios (using the 10\,\% normalization, 
see Appendix) for the new sample, are given in Table~{\ref{tab:kinematics}.

%-----------------------------------------------------------------------------
\begin{table}
\caption{
        Wavelength coverage for the different spectral 
        orders (SO) for the two settings (Blue and Red) of the CCD.
        }
\centering \scriptsize
\begin{tabular}{ccc|ccc}
\hline \hline\noalign{\smallskip}
SO      &  Blue
        &  Red
        &  SO
        &  Blue
        &  Red     \\
        &  [{\AA}]
        &  [{\AA}]
        & 
        &  [{\AA}]
        &  [{\AA}] \\
\noalign{\smallskip}
\hline\noalign{\smallskip}
50 &                & 4530 - 4567  &  37 & 6030 - 6084    & 6121 - 6172 \\
49 & 4552 - 4594    & 4622 - 4660  &  36 & 6198 - 6253    & 6291 - 6344 \\
48 & 4648 - 4690    & 4718 - 4758  &  35 & 6375 - 6430    & 6471 - 6524 \\
47 & 4748 - 4789    & 4819 - 4858  &  34 & 6562 - 6620    & 6660 - 6716 \\
46 & 4850 - 4893    & 4924 - 4964  &  33 & 6760 - 6820    & 6863 - 6920 \\
45 & 4958 - 5002    & 5035 - 5074  &  32 & 6972 - 7035    & 7077 - 7136 \\
44 & 5072 - 5116    & 5148 - 5190  &  31 & 7197 - 7260    & 7305 - 7366 \\
43 & 5190 - 5235    & 5267 - 5310  &  30 & 7437 - 7503    & 7550 - 7612 \\
42 & 5312 - 5360    & 5392 - 5437  &  29 & 7695 - 7762    & 7810 - 7875 \\
41 & 5442 - 5490    & 5526 - 5570  &  28 & 7968 - 8040    & 8088 - 8155 \\
40 & 5578 - 5628    & 5662 - 5709  &  27 & 8265 - 8337    & 8388 - 8457 \\
39 & 5722 - 5772    & 5807 - 5855  &  26 & 8582 - 8657    & 8710 - 8783 \\
38 & 5872 - 5924    & 5960 - 6010  &     &                &             \\
\noalign{\smallskip}
\hline
\label{tab:sofin}
\end{tabular}
\end{table}
%----------------------------------------------------------------------------

%----------------------------------------------------------------------------
\begin{table*}[ht]
\centering
\caption{
        Our program stars. 
        Columns~1--3 give the identifications for each star,
        Hipparcos, HD, and HR numbers; col.~4 gives the spectral 
        class as listed in the SIMBAD database; cols.~5--7 give $V$ magnitude, 
        parallax ($\pi$), and accuracy of the parallax ($\sigma_{\pi}$), 
        all from the Hipparcos catalogue; cols.~8--10 give the 
        stellar atmospheric parameters, metallicity ([Fe/H]), 
        effective temperature ($T_{\rm eff}$), and surface gravity ($\log g$); 
        col.~11 gives the microturbulence ($\xi_{\rm t}$); col.~12 gives the 
        stellar mass ($\mathcal{M}$); col.~13 the bolometric correction 
        ($BV$). The last column indicates which instrument was used to obtain
        the spectrum.
        }
\centering %\scriptsize%\tiny
\begin{tabular}{rrrlcccrccccrl}
\hline \hline\noalign{\smallskip}
           \multicolumn{3}{c}{Identifications}
        &  Sp. type
        &  $V$
        &  \multicolumn{1}{c}{$\pi$}
        &  \multicolumn{1}{c}{$\sigma_{\pi}$}
        &  [Fe/H]
        &  $T_{\rm eff}$
        &  $\log g$ 
        &  $\xi_{\rm t}$
        &  \multicolumn{1}{c}{$\mathcal{M}$}
        &  $BC$
        &  Spec. \\
\noalign{\smallskip}
           \multicolumn{1}{c}{HIP} 
        &  \multicolumn{1}{c}{HD} 
        &  \multicolumn{1}{c}{HR} 
        & 
        &  [mag] 
        &  [mas] 
        &  [mas]
        &   
        &  [K] 
        &  [cgs] 
        &  [$\rm km\,s^{-1}$] 
        &  [M$_{\sun}$]
        &  [mag] 
        & \\
\noalign{\smallskip}                      
\hline
\noalign{\smallskip}
\multicolumn{14}{c}{\bf THIN DISK STARS} \\
\noalign{\smallskip}
    699 &    400 &   17 & F8IV       & 6.21 &  30.26 &  0.69 &  $-0.20$  & 6250 &  4.19  &  1.35  & 1.27 &  $-0.09$ & SOFIN  \\
    910 &    693 &   33 & F5V        & 4.89 &  52.94 &  0.77 &  $-0.36$  & 6220 &  4.07  &  1.43  & 1.10 &  $-0.10$ & SOFIN  \\
   2235 &   2454 &  107 & F6V        & 6.05 &  27.51 &  0.86 &  $-0.28$  & 6645 &  4.17  &  1.75  & 1.30 &  $-0.07$ & SOFIN  \\
   2787 &   3229 &  143 & F5IV       & 5.94 &  17.97 &  0.74 &  $-0.11$  & 6620 &  3.86  &  1.70  & 1.68 &  $-0.06$ & SOFIN  \\
   3909 &   4813 &  235 & F7IV       & 5.17 &  64.69 &  1.03 &  $-0.06$  & 6270 &  4.41  &  1.12  & 1.17 &  $-0.08$ & SOFIN  \\
   5862 &   7570 &  370 & F8V        & 4.97 &  66.43 &  0.64 &  $ 0.17$  & 6100 &  4.26  &  1.10  & 1.04 &  $-0.08$ & UVES   \\
  10306 &  13555 &  646 & F5V        & 5.23 &  33.19 &  0.85 &  $-0.17$  & 6560 &  4.04  &  1.75  & 1.48 &  $-0.07$ & SOFIN  \\
  15131 &  20407 &      & G1V        & 6.75 &  41.05 &  0.59 &  $-0.52$  & 5834 &  4.35  &  1.00  & 0.85 &  $-0.15$ & UVES   \\
  18833 &  25322 &      & F5V        & 7.82 &  11.47 &  1.02 &  $-0.52$  & 6370 &  3.99  &  1.75  & 1.19 &  $-0.10$ & SOFIN  \\
  88945 & 166435 &      & G0         & 6.84 &  39.62 &  0.68 &  $-0.05$  & 5690 &  4.37  &  1.33  & 0.95 &  $-0.15$ & SOFIN  \\
  92270 & 174160 & 7079 & F8V        & 6.19 &  34.85 &  0.68 &  $-0.06$  & 6370 &  4.32  &  1.50  & 1.20 &  $-0.07$ & SOFIN  \\
  93185 & 176377 &      & G0         & 6.80 &  42.68 &  0.64 &  $-0.28$  & 5810 &  4.40  &  0.90  & 0.84 &  $-0.14$ & SOFIN  \\
  96258 & 184960 & 7451 & F7V        & 5.71 &  39.08 &  0.47 &  $-0.02$  & 6380 &  4.25  &  1.48  & 1.25 &  $-0.08$ & SOFIN  \\
 107975 & 207978 & 8354 & F6IV       & 5.52 &  36.15 &  0.69 &  $-0.53$  & 6460 &  4.06  &  1.50  & 1.10 &  $-0.10$ & SOFIN  \\
 113174 & 216756 & 8718 & F5II       & 5.91 &  24.24 &  0.68 &  $-0.11$  & 6870 &  4.14  &  1.95  & 1.51 &  $-0.05$ & SOFIN  \\
\noalign{\smallskip}
\hline
\noalign{\smallskip}
\multicolumn{14}{c}{\bf THICK DISK STARS} \\
\noalign{\smallskip}
  11309 &  15029 &      & F5         & 7.36 &  15.05 &  0.91 &  $-0.32$  & 6210 &  3.98  &  1.40  & 1.12 &  $-0.10$ & SOFIN  \\
  12306 &  16397 &      & G0V        & 7.36 &  27.89 &  1.12 &  $-0.53$  & 5765 &  4.20  &  0.90  & 0.78 &  $-0.16$ & SOFIN  \\
  15510 &  20794 & 1008 & G8V        & 4.26 & 165.02 &  0.55 &  $-0.41$  & 5480 &  4.43  &  0.75  & 0.82 &  $-0.20$ & UVES   \\
  16788 &  22309 &      & G0         & 7.65 &  22.25 &  1.14 &  $-0.32$  & 5920 &  4.24  &  1.00  & 0.89 &  $-0.13$ & SOFIN  \\
  18235 &  24616 &      & G8IV       & 6.68 &  15.87 &  0.81 &  $-0.71$  & 5000 &  3.13  &  0.95  & 0.79 &  $-0.33$ & SOFIN  \\
  20242 &  27485 &      & G0         & 7.87 &  14.79 &  0.98 &  $-0.26$  & 5650 &  3.94  &  1.00  & 1.02 &  $-0.16$ & SOFIN  \\
  21832 &  29587 &      & G2V        & 7.29 &  35.31 &  1.07 &  $-0.61$  & 5570 &  4.27  &  0.65  & 0.72 &  $-0.19$ & SOFIN  \\
  26828 &  37739 &      & F5         & 7.92 &  12.32 &  0.99 &  $-0.37$  & 6410 &  4.15  &  1.50  & 1.31 &  $-0.09$ & SOFIN  \\
  36874 &  60298 &      & G2V        & 7.37 &  25.42 &  0.99 &  $-0.07$  & 5730 &  4.22  &  0.90  & 0.98 &  $-0.14$ & SOFIN  \\
  37789 &  62301 &      & F8V        & 6.74 &  29.22 &  0.96 &  $-0.67$  & 5900 &  4.09  &  1.20  & 0.88 &  $-0.15$ & SOFIN  \\
  40613 &  69611 &      & F8         & 7.74 &  20.46 &  1.16 &  $-0.63$  & 5740 &  4.11  &  0.92  & 0.84 &  $-0.16$ & SOFIN  \\
  44075 &  76932 & 3578 & F8IV       & 5.80 &  46.90 &  0.97 &  $-0.91$  & 5875 &  4.10  &  1.10  & 0.86 &  $-0.17$ & SOFIN  \\
  44860 &  78558 &      & G3V        & 7.29 &  27.27 &  0.91 &  $-0.45$  & 5690 &  4.19  &  0.82  & 0.90 &  $-0.16$ & SOFIN  \\
 112151 & 215110 &      & G5         & 7.73 &  11.30 &  0.93 &  $-0.42$  & 5035 &  3.43  &  0.85  & 1.15 &  $-0.32$ & SOFIN  \\
 116421 & 221830 &      & F9V        & 6.86 &  30.93 &  0.73 &  $-0.45$  & 5700 &  4.15  &  0.95  & 0.93 &  $-0.16$ & SOFIN  \\
 116740 & 222317 &      & G2V        & 7.04 &  20.27 &  0.76 &  $ 0.05$  & 5740 &  3.96  &  1.15  & 1.12 &  $-0.13$ & SOFIN  \\
 118010 & 224233 &      & G0         & 7.67 &  20.01 &  0.74 &  $-0.07$  & 5795 &  4.17  &  1.00  & 1.03 &  $-0.13$ & SOFIN  \\
\noalign{\smallskip}
\hline
\noalign{\smallskip}
\multicolumn{14}{c}{\bf ``TRANSITION OBJECTS"} \\
\noalign{\smallskip}
   3170 &   3823 &  176 & G1V        & 5.89 &  39.26 &  0.56 &  $-0.34$  & 5970 &  4.11  &  1.40  & 1.05 &  $-0.12$ & FEROS  \\
  44441 &  77408 &      & F6IV       & 7.03 &  19.85 &  0.86 &  $-0.28$  & 6260 &  4.11  &  1.48  & 1.13 &  $-0.09$ & SOFIN  \\
  95447 & 182572 & 7373 & G8IV       & 5.17 &  66.01 &  0.77 &  $ 0.37$  & 5600 &  4.13  &  1.10  & 0.98 &  $-0.14$ & FEROS  \\
 100412 & 193307 & 7766 & G0V        & 6.26 &  30.84 &  0.88 &  $-0.32$  & 5960 &  4.06  &  1.20  & 1.07 &  $-0.12$ & FEROS  \\
\noalign{\smallskip}                  
\hline
\end{tabular}
\label{tab:parameters}
\end{table*}
%------------------------------------------------------------------------------

%==============================================================================
\section{Observations and data reductions}  \label{sec:sofin}

%==============================================================================
\subsection{SOFIN data}

Observations were carried out with the Nordic Optical Telescope (NOT)
on La Palma, Spain, during two observing runs in August (3 nights) and
November (5 nights) 2002. The SOFIN (SOviet FINnish) spectrograph was
used to obtain spectra with high resolving power ($R \sim 80\,000$)
and high signal-to-noise ratios ($S/N \gtrsim 250$). A solar spectrum
was also obtained by observing the Moon. To avoid long exposure times
and thus the effects of cosmic rays the exposures were split
into two or three exposures (not longer than $\sim 20$\,min).

The spectra were reduced using the 4A package (Ilyin~\cite{ilyin}).
This comprises a standard procedure for data reduction and includes
bias subtraction, estimation of the variances of the pixel
intensities, correction for the master flat field, scattered light
subtraction with the aid of 2D-smoothing splines, definition of the
spectral orders, and weighted integration of the intensity with
elimination of cosmic rays. The wavelength calibration was done using
ThAr comparison spectra, one taken before and one after each
individual object exposure.  A typical error of the ThAr wavelength
calibration is about 10\,m\,s$^{-1}$ in the image center.

In order to get large enough spectral wavelength coverage we observed
each star twice with different settings for the CCD (see
Table~\ref{tab:sofin}).  Each setting resulted in $\sim 45$ spectral
orders. In this study we use spectral lines with wavelengths ranging
from $\sim 4500$\,{\AA} to $\sim 8800$\,{\AA}, i.e., spectral orders 
26--50.  For each star we analyzed approximately 260 spectral lines, 
which form a sub-set of the 450 lines that were analyzed for each star 
in Bensby et al.~(\cite{bensby}).

In total we observed 41 stars with NOT/SOFIN, but unfortunately we had
to reject 11 of them from the analysis because their rotational
velocities ($v\cdot \sin i$) were too high to allow equivalent width
measurements (HIP~3641, HIP~4989, HIP~5034, HIP~6669, HIP~6706,
HIP~18859, HIP~24109, HIP~45879, and HIP~87958) or because they were
found to be spectroscopic binaries (HIP~17732 and HIP~109652).

%==============================================================================
\subsection{FEROS and UVES data}

Spectra for 69 stars (abundances for 66 of these were presented in
Bensby et al.~\cite{bensby}, the remaining three
will be discussed here and are labeled as ``transition
objects") were obtained with the FEROS spectrograph
on the ESO 1.52-m telescope on La Silla in Chile in September 2000 and
August/September 2001.  These spectra have $R \sim 48\,000$ and $S/N
\sim 150$--250 with a wavelength coverage that is complete from
$\sim4000$\,{\AA} to $\sim9400$\,{\AA}. In each stellar spectrum we
analyze a total of $\sim 450$ spectral lines. The reductions and the
analysis of these stars were presented in Bensby et
al.~(\cite{bensby}).

Spectra for three additional stars were obtained with the UVES
spectrograph on the VLT/Keuyen 8-m telescope in July 2002. The spectra
have $R \sim 110\,000$ and $S/N \gtrsim 350$. The reductions of these
observations will be discussed in a forthcoming paper where the
majority of the stars (bulge and thick disk {\it in situ} giants) from
that observing run will be presented. The setting of the CCD gives a
wavelength coverage from $\sim 5540$\,{\AA} to $\sim 7560$\,{\AA}
(with a gap between 6520--6670\,{\AA}). This resulted in that $\sim
200$ spectral lines were analyzed (again a subset of the 450 lines
analyzed in Bensby et al.~\cite{bensby}).

%==============================================================================
\section{Stellar model atmospheres}  \label{sec:atmospheres}

%---------------------------------------------------------------------------
\begin{figure}
\resizebox{\hsize}{!}{
        \includegraphics[bb=18 144 592 375,clip]{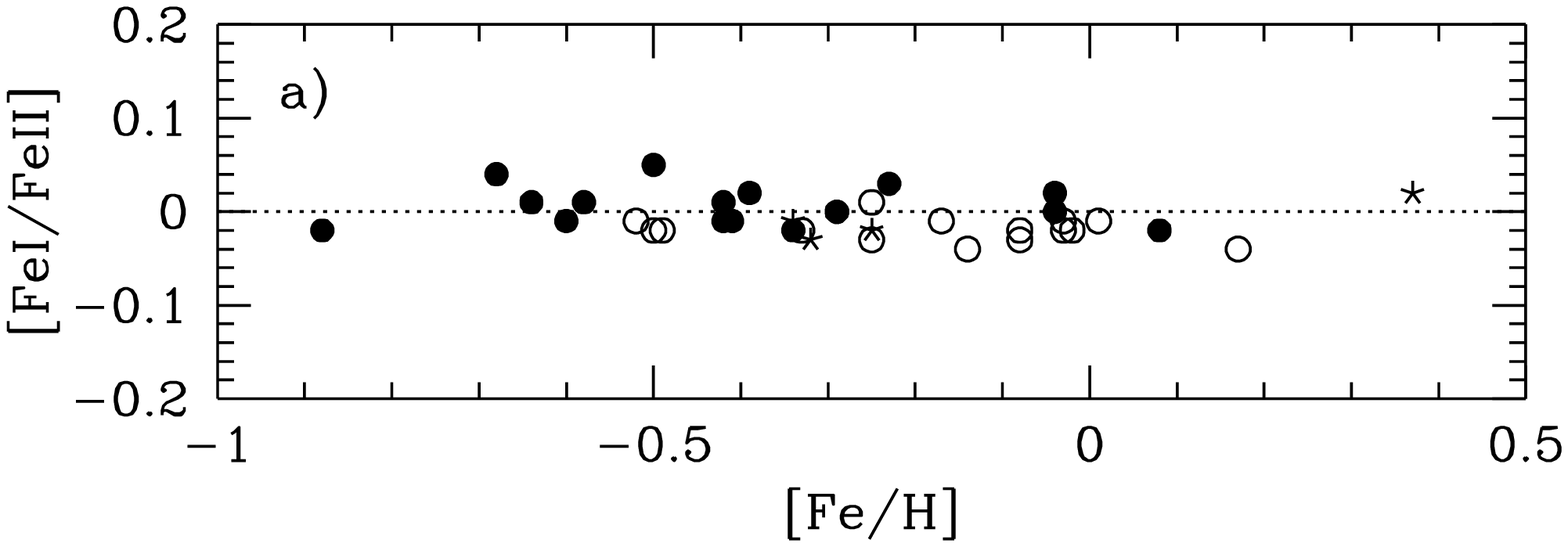}}
\resizebox{\hsize}{!}{
        \includegraphics[bb=18 144 592 350,clip]{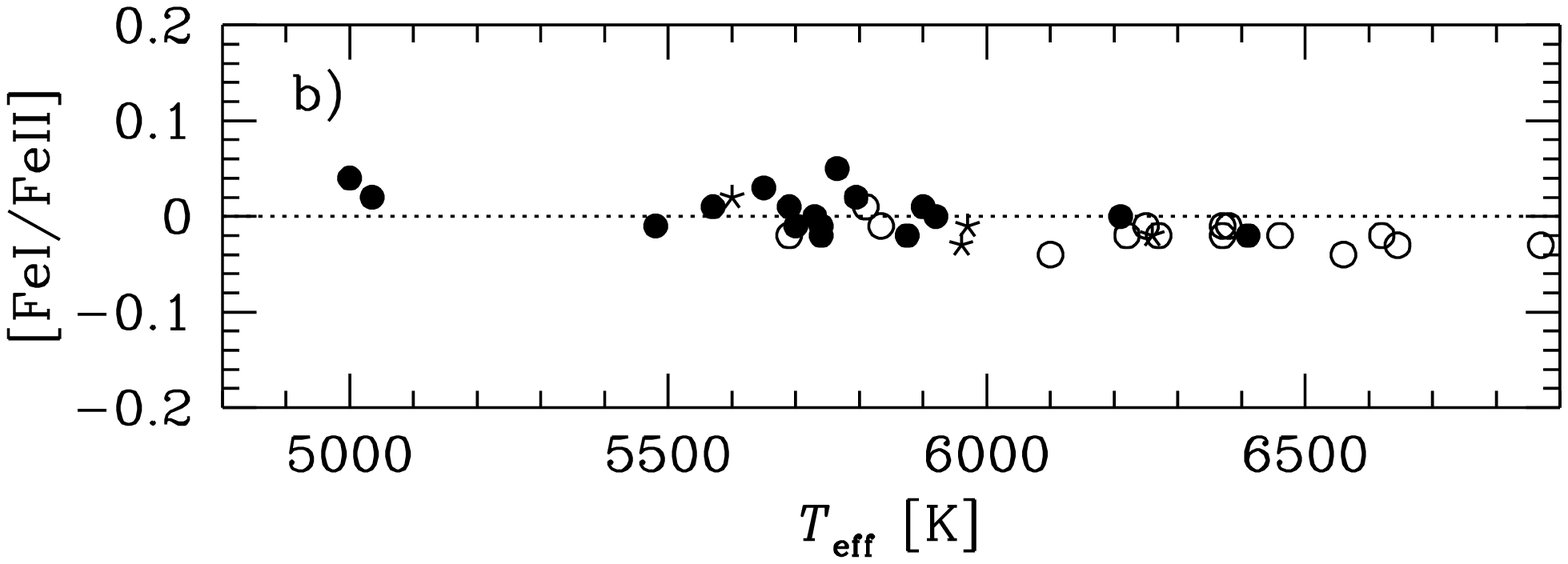}}
\resizebox{\hsize}{!}{
        \includegraphics[bb=18 144 592 350,clip]{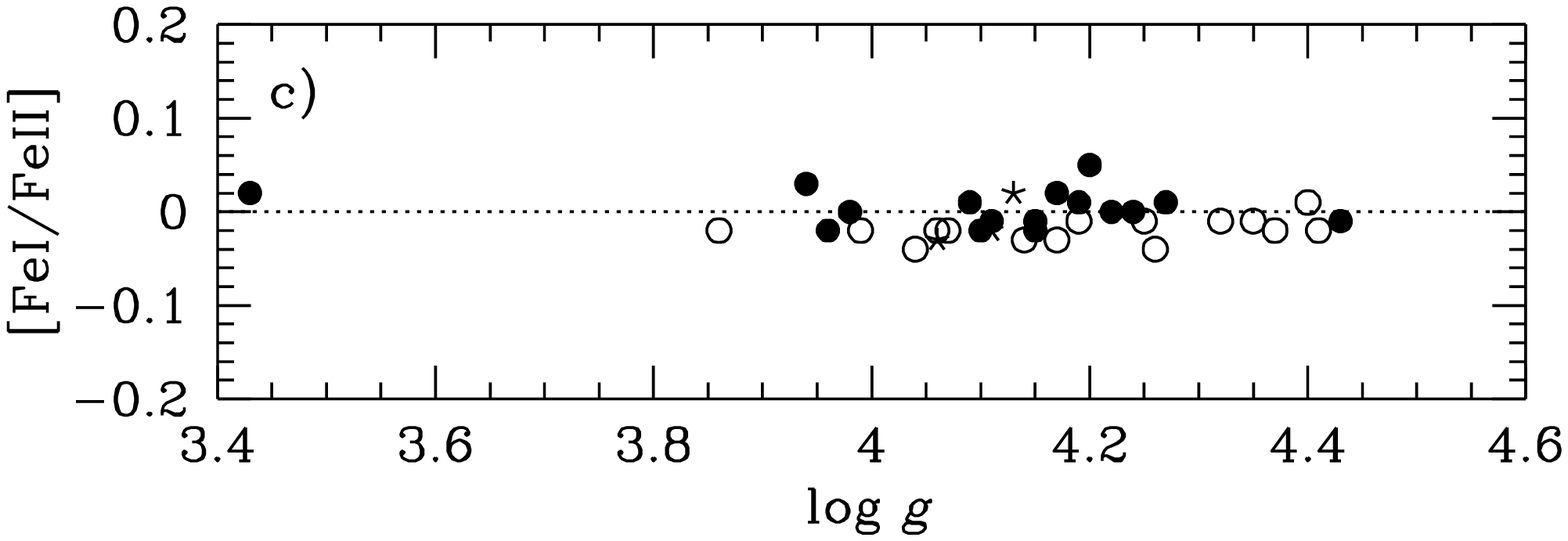}}
\resizebox{\hsize}{!}{
        \includegraphics[bb=18 144 592 350,clip]{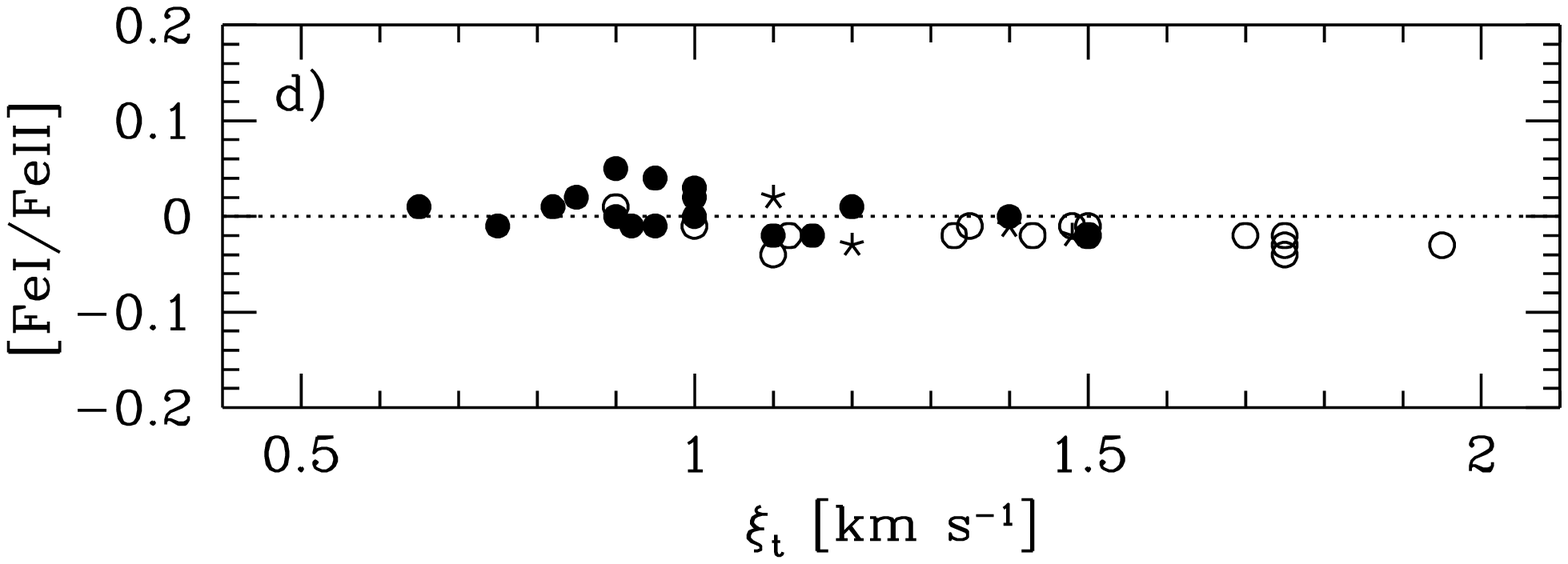}}
\caption{
        $\rm [\ion{Fe}{i}/\ion{Fe}{ii}]$ versus $\rm [Fe/H]$,
        $\log T_{\rm eff}$, $\log g$, and $\xi_{\rm t}$.
        Thin and thick disk stars are marked by open and filled circles,
        respectively, and transition objects by ``stars".
        }
\label{fig:fediff}
\end{figure}
%--------------------------------------------------------------------------
The calculation of stellar atmospheric models were done with the Uppsala 
MARCS code (Gustafsson et al.~\cite{gustafsson};
Edvardsson et al.~\cite{edvardsson}; Asplund et al.~\cite{asplund}). 
The iterative process to tune the stellar parameters used in the 
construction of the model atmospheres and the abundance analysis
is fully described in Bensby et al.~(\cite{bensby}). In summary the 
main ingredients are: Effective temperature ($T_{\rm eff}$) is determined 
by requiring \ion{Fe}{i} lines with different lower excitation potentials 
to give equal abundances; Stellar mass ($\mathcal{M}$) is estimated from 
the evolutionary tracks by Yi et al.~(\cite{yi2}). These two parameters are 
then used together with parallaxes and magnitudes from the Hipparcos 
catalogue (ESA~\cite{esa}) to determine the surface gravity ($\log g$) of 
the star (see Eq.~4 in Bensby et al.~\cite{bensby}). The bolometric 
correction ($BC$) which also is needed to determine $\log g$ is found by 
interpolating in the grids by 
Alonso et al.~(\cite{alonso}). The microturbulence parameter ($\xi_{\rm t}$) 
is determined by forcing all \ion{Fe}{i} lines to give the same abundance 
regardless of line strength, i.e. $\log (W_{\lambda}/\lambda)$.

Since the \ion{Fe}{ii} lines have not been used in the determination of the
atmospheric parameters they can be used to check the derived parameters
as well as the \ion{Fe}{i} abundances. This is an important test since the
\ion{Fe}{i} abundancs can be affected by NLTE effects while \ion{Fe}{ii} lines 
generally are not (e.g. Th\'evenin \& Idiart~\cite{thevenin}; 
Gratton et al.~\cite{gratton}). Figure~\ref{fig:fediff} shows the difference
[\ion{Fe}{i}/\ion{Fe}{ii}] versus [Fe/H], $T_{\rm eff}$, $\log g$, and 
$\xi_{\rm t}$. There are slight indications that the \ion{Fe}{i} abundances
come out too low for the stars with the highest $T_{\rm eff}$ and $\xi_{\rm t}$
(Figs.~\ref{fig:fediff}b and d). The effect seems to be small ($<0.05$\,dex)
and the number of stars at these higher values of $T_{\rm eff}$ and 
$\xi_{\rm t}$ are too low to allow us to delve deeper into this.
Generally there are no significant differences or trends with either of
the atmospheric parameters, which indicates that NLTE effects for \ion{Fe}{i}
are not severe for the majority of our stars.
This is also true for the stars analyzed in Bensby et al.~(\cite{bensby}).
The derived stellar parameters are listed in Table~\ref{tab:parameters}
for the 36 new stars. The parameters for the remaining
66 stars are given in Bensby et al.~(\cite{bensby}).

%==============================================================================
\section{Abundance analysis}  \label{sec:analysis}

\subsection{Methods and solar analysis} \label{sec:solaranalysis}

\subsubsection{Abundances from equivalent widths}

%------------------------------------------------------------------------------
\begin{figure}
 \resizebox{\hsize}{!}{\includegraphics[bb=18 144 592 470,clip]{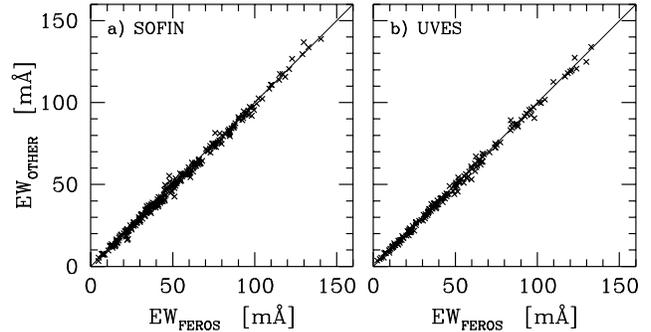}}
\caption{
        Comparison between equivalent widths measured in the FEROS solar
        spectrum and the SOFIN and UVES solar spectra.
        The average differences are
        $\rm \langle EW_{FEROS}-EW_{SOFIN}\rangle = 1.02\pm 2.1$\,m{\AA}
        (251 lines in common) and
        $\rm \langle EW_{FEROS}-EW_{UVES}\rangle = 0.58\pm 1.9$\,m{\AA}.
        (165 lines in common)
         }
\label{fig:ew_sof_fer}
\end{figure}
%------------------------------------------------------------------------------

%------------------------------------------------------------------------------
\begin{figure*} 
 \resizebox{\hsize}{!}{\includegraphics[bb=18 144 592 618,clip]{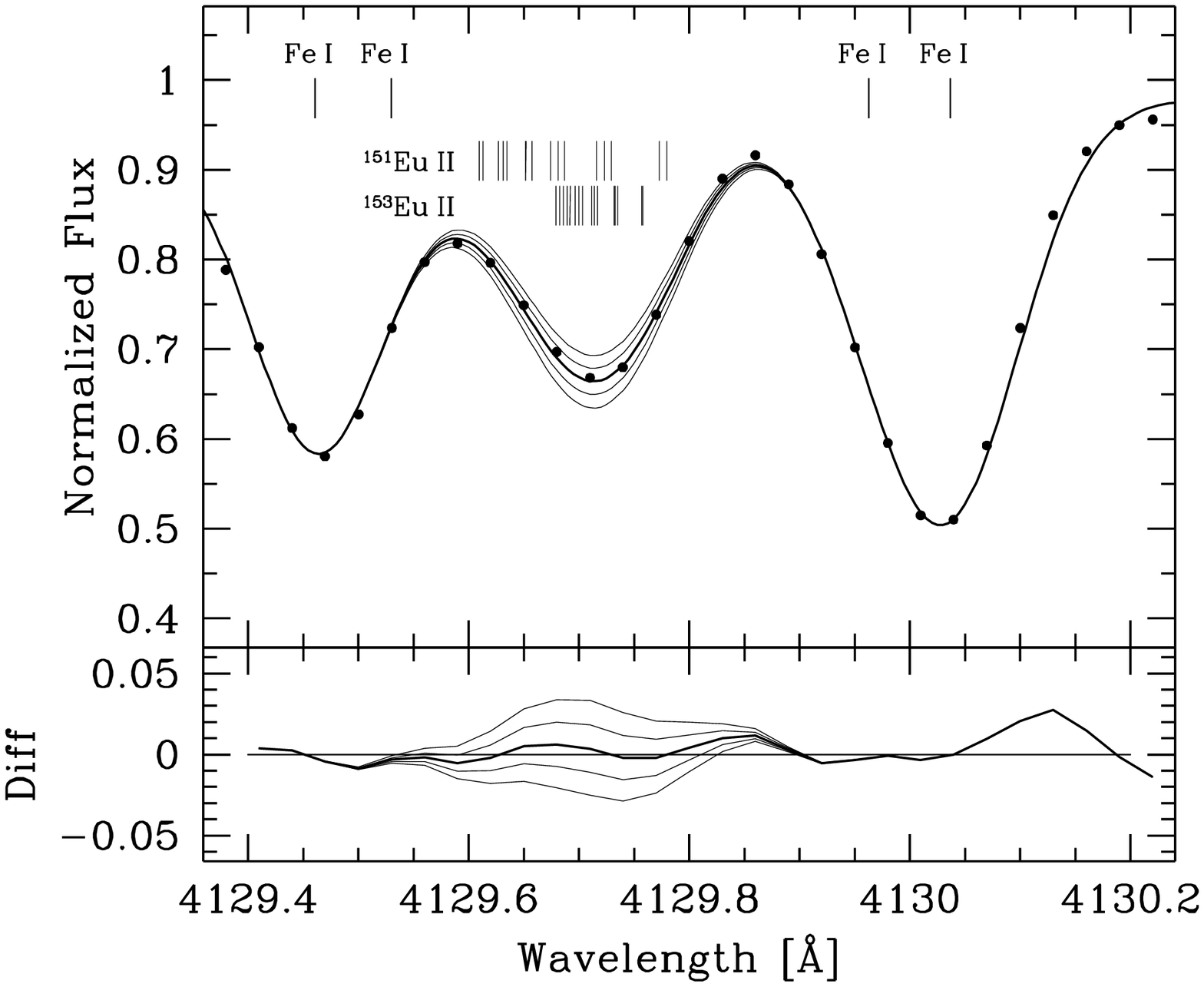}
                       \includegraphics[bb=18 144 592 618,clip]{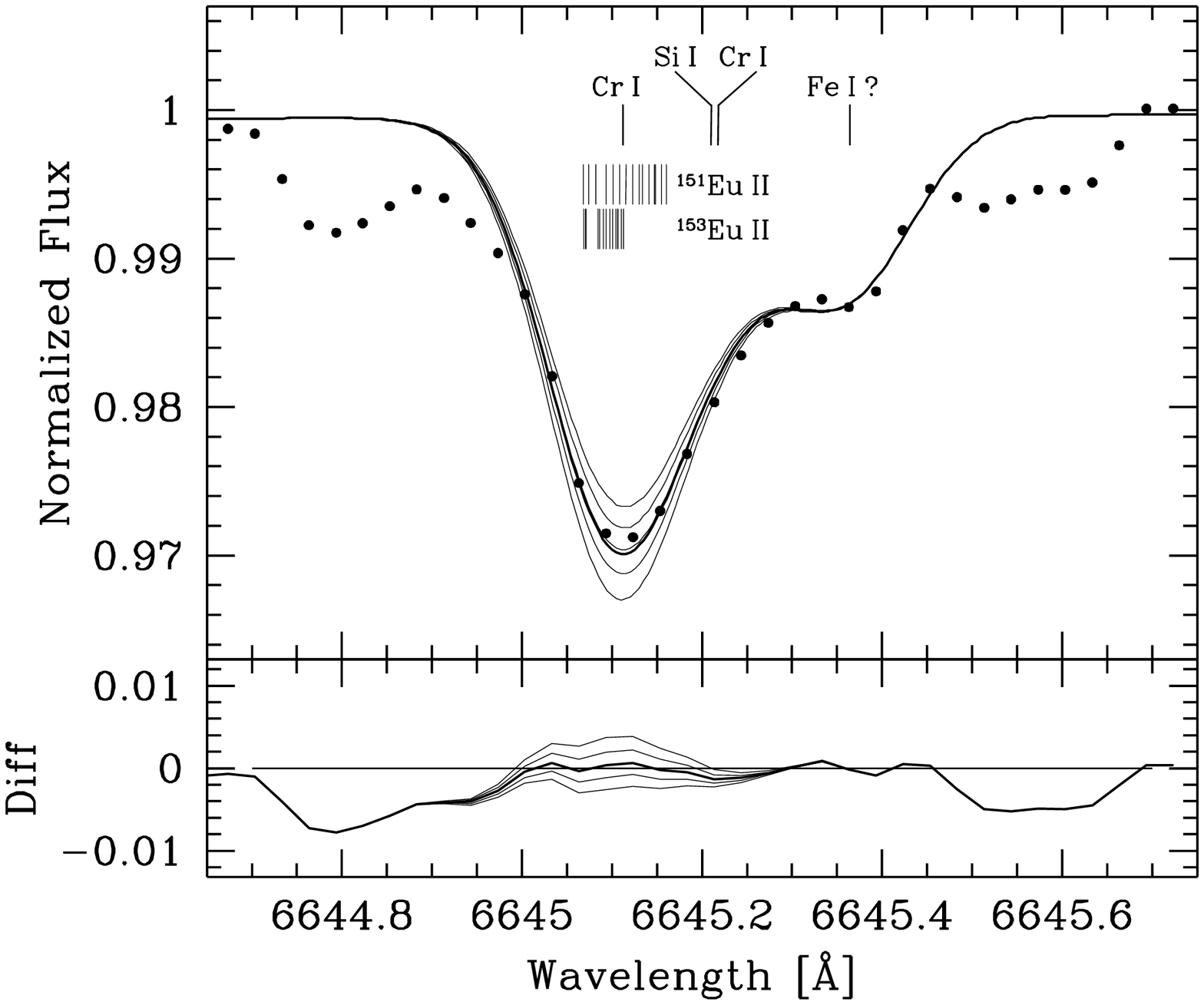}}
\caption{
        Synthetic and observed solar (FEROS) spectra for: {\bf a)}
        the \ion{Eu}{ii} at 4129\,{\AA}, and {\bf b)} the \ion{Eu}{ii}
        line at 6645\,{\AA}. Five synthetic spectra with different Eu 
        abundances, in steps of 0.03\,dex have been plotted for each line. 
        The hyperfine components for the two Eu isotopes are indicated as 
        well as other important lines in the regions. The bottom panels shows 
        the differences between the observed and synthetic spectra.
         }
\label{fig:eu_synt}
\end{figure*}
%------------------------------------------------------------------------------
%------------------------------------------------------------------------------
\begin{table}[b]
\caption{
        Atomic line data. 
        Column~1 gives the element;
        col.~2 the wavelength,
        col.~3 the lower excitation potential;
        col.~4 the correction factor to the classical Uns\"old damping 
        constant;
        col.~5 indicates if the broadening by collisions have been taken from
        Anstee et al.~(\cite{anstee}), 
        Barklem \& O'Mara~(\cite{barklem1},\cite{barklem3}), and 
        Barklem et al.~(\cite{barklem2}, \cite{barklem4})
        (indicated by an ``S") instead of the classical Uns\"old broadening
        (indicated by an ``U"). Column~6 gives the radiation
        damping constant;
        col.~7 gives the $\log gf$-values;
        col.~8 gives the references for the $\log gf$-values.
        The full
        table is available in electronic form at the CDS via anonymous ftp to
        \texttt{cdsarc.u-strasbg.fr} (130.79.128.5) or via
        \texttt{cdsweb.u-strasbg.fr/cats/cgi-bin/qcat?J/A+A/}
        }
\centering %\scriptsize
\setlength{\tabcolsep}{2.5mm}
\begin{tabular}{lccccrc}
\hline \hline\noalign{\smallskip}
           $\lambda$
        &  $\chi_{\rm l}$
        &  $\delta \gamma_{6}$
        &  DMP
        &  $\gamma_{\rm rad}$
        &  $\log gf$
        &  Ref.  \\
\noalign{\smallskip}
           (\AA )
        &  (eV)
        &
        &
        &  (s$^{-1}$)
        &
        &        \\
\noalign{\smallskip}
\hline\noalign{\smallskip}
      \multicolumn{7}{l}{{\bf Y\,{\sc ii}} $\log \epsilon_{\odot} = 2.24$} \\
     4854.87  &  0.99   &  2.50 & U &  1.0e+08 & $-0.11$ &  PN   \\
     4883.68  &  1.08   &  2.50 & U &  1.0e+08 & $ 0.23$ &  PN   \\
  \vdots  & \vdots  & \vdots & \vdots & \vdots & \vdots  &  \vdots  \\
\noalign{\smallskip}
\hline
\end{tabular}
\label{tab:atomdata}
\end{table}
%--------------------------------------------------------------------------
Elemental abundances for Na, Mg, Al, Si, Ca, Ti, Cr, Fe, Ni, Zn, Y,
and Ba have been determined by means of equivalent width
measurements. The method and the atomic data are the same as in Bensby
et al.~(\cite{bensby}) and are extensively described therein, except
for the \ion{Y}{ii} and \ion{Ba}{ii} lines that are listed in
Table~\ref{tab:atomdata}. The $\log gf$-values for these lines were
taken from Pitts \& Newsom~(\cite{pitts}) for \ion{Y}{ii}, and Sneden
et al.~(\cite{sneden}) and Gallagher~(\cite{gallagher}) for
\ion{Ba}{ii}.  We have not taken hyperfine structure or isotopic
shifts into account when measuring equivalent widths for the
\ion{Ba}{ii} lines.  The structure of the \ion{Ba}{ii} lines that we
have used are dominated by strong central peaks, containing the even
isotopes, and smaller peaks on the sides caused by the hyperfine
components of the odd isotopes (Karlsson \&
Litz\'en~\cite{karlsson}). The shifts of the even isotopes in the
central peaks are indeed small, and are not resolvable in our
spectra. Hence the \ion{Ba}{ii} lines have essentially Gaussian line
profiles.

Figure~\ref{fig:ew_sof_fer}a shows a comparison between the equivalent
width measurements in the FEROS solar spectrum and the SOFIN solar
spectrum. For the 251 lines in common between the two spectra there is
a slight offset present. The FEROS equivalent widths are on average
$1.02\pm 2.1$\,m{\AA} larger.  There is no obvious reason for this. It
can, however, be due to the different resolutions of the two
spectrographs. In the SOFIN spectra, with their higher resolution, it
is easier to avoid small blends that are not detected in the FEROS
spectra, and therefore the equivalent width could be on average
smaller. The SOFIN spectra in general also have higher $S/N$ ratios
which could lead to a lower placement of the continuum, as compared to
the more noisy spectra from FEROS, when doing the measurements.
Figure~\ref{fig:ew_sof_fer}b shows a similar comparison between the
FEROS and the UVES\footnote{We did not obtain a solar spectrum with
UVES during our observing run. Instead we have used the UVES solar
spectrum that is available on the web: {\tt www.eso.org/observing/
dfo/quality/UVES/pipeline/solar\_spectrum.html}} solar spectra. The
difference is smaller than between SOFIN and FEROS, but the trend
persists, i.e. that the FEROS equivalent widths are slightly higher.
          
The derived solar elemental abundances are tabulated in
Table~\ref{tab:solarvalues} for the different spectrographs.  As
expected the SOFIN abundances are somewhat lower.  In order to put all
observations on a common baseline we subtract the difference between
the solar abundances we derive and the standard photospheric
abundances as given in Grevesse \& Sauval~(\cite{grevesse}).  It
should be emphasized that this normalization is done individually for
each set of abundances for the different spectrographs.

%------------------------------------------------------------------------------
\begin{table*}
\caption{
        Elemental abundances from the solar analysis. The first column gives 
        the element and degree of ionization. An asterisk in the second
        column indicates that the $\log gf$-values for these lines are 
        astrophysical (originating from Bensby et al.~\cite{bensby}), and an 
        asterisk within parenthesis that a part of the lines have 
        astrophysical $\log gf$-values. 
        In the lower part of the table, col.~2 is used to indicate the
        wavelength of the spectral line.
        The third column gives the standard 
        solar photospheric abundance as given in 
        Grevesse \& Sauval~(\cite{grevesse}). Columns 4\,--\,6 give our
        solar analysis based on the FEROS spectra 
        (see Bensby et al.~\cite{bensby}), columns 7\,--\,9 our solar 
        analysis based on the SOFIN spectra, and columns 10\,--\,12 our
        solar analysis based on the UVES spectra. For each study we give the
        number of lines that were analyzed ($N_{\rm lines}$), the mean 
        abundance from these lines ($\epsilon(X)$), and the difference 
        (Diff.) compared to the standard photospheric value in column~3. 
        }
\centering %\scriptsize
\begin{tabular}{lccrcrrcrrcr}
\hline \hline\noalign{\smallskip}
Ion     &  
        &  Phot.
        &  \multicolumn{3}{c}{FEROS/CES}
        &  \multicolumn{3}{c}{SOFIN} 
        &  \multicolumn{3}{c}{UVES} \\
\noalign{\smallskip}
        &
        &  $\epsilon(X)$
        &  $N_{\rm lin}$
        &  $\epsilon(X)$
        &  Diff.
        &  $N_{\rm lin}$             
        &  $\epsilon(X)$
        &  Diff.          
        &  $N_{\rm lin}$
        &  $\epsilon(X)$
        &  Diff.  \\
\noalign{\smallskip}
\hline\noalign{\smallskip}
\ion{Fe}{i}  &     &  7.50  & 147    &  7.56  & $+$0.06 &  86    &  7.53  & $+$0.03 &   62   &  7.56  & $+$0.06       \\
\ion{Fe}{ii} &     &  7.50  & 29     &  7.58  & $+$0.08 &  19    &  7.53  & $+$0.03 &   13   &  7.59  & $+$0.09       \\
\ion{Na}{i}  &     &  6.33  & 4      &  6.27  & $-$0.06 &   4    &  6.27  & $-$0.06 &    4   &  6.25  & $-$0.08       \\
\ion{Mg}{i}  &  *  &  7.58  & 7      &  7.58  &    0.00 &   4    &  7.57  & $-$0.01 &    1   &  7.58  &    0.00       \\
\ion{Al}{i}  &  *  &  6.47  & 6      &  6.47  &    0.00 &   6    &  6.47  &    0.00 &    2   &  6.50  & $+$0.03       \\
\ion{Si}{i}  & (*) &  7.55  & 32     &  7.54  & $-$0.01 &  15    &  7.54  & $-$0.02 &   18   &  7.54  & $-$0.01       \\
\ion{Ca}{i}  &  *  &  6.36  & 22     &  6.36  &    0.00 &  10    &  6.35  & $-$0.01 &   12   &  6.37  & $+$0.01       \\
\ion{Ti}{i}  &     &  5.02  & 31     &  4.92  & $-$0.10 &  12    &  4.90  & $-$0.12 &    7   &  4.93  & $-$0.09       \\
\ion{Ti}{ii} &     &  5.02  & 18     &  4.91  & $-$0.11 &   8    &  4.88  & $-$0.11 &    0   &        &               \\
\ion{Cr}{i}  &  *  &  5.67  & 14     &  5.67  &    0.00 &   6    &  5.64  & $-$0.03 &    2   &  5.76  & $+$0.09       \\
\ion{Cr}{ii} &  *  &  5.67  & 9      &  5.67  &    0.00 &   6    &  5.60  & $-$0.07 &    0   &        &               \\
\ion{Ni}{i}  & (*) &  6.25  & 54     &  6.25  & $-$0.01 &  36    &  6.24  & $-$0.01 &   23   &  6.24  & $-$0.01       \\
\ion{Zn}{i}  &  *  &  4.60  & 2      &  4.60  &    0.00 &   1    &  4.58  & $-$0.02 &    1   &  4.55  & $-$0.05       \\
\ion{Y}{ii}  &     &  2.24  & 7      &  2.20  & $-$0.04 &   4    &  2.05  & $-$0.19 &    0   &        &               \\
\ion{Ba}{ii} &     &  2.13  & 4      &  2.29  & $+$0.16 &   4    &  2.27  & $+$0.14 &    3   &  2.34  & $+$0.21       \\
\noalign{\smallskip}
\hline
\noalign{\smallskip}
$\rm [\ion{O}{i}]$ & 6300 & 8.83  &  &  8.71  & $-$0.12 &        &  8.74  &  $-$0.09&        &        &        \\
$\rm [\ion{O}{i}]$ & 6363 & 8.83  &  &  9.06  & $+$0.23 &        &        &         &        &        &        \\
\ion{O}{i}   & 7771 & 8.83  &        &  8.83  & $\pm$0.00 &      &        &         &        &        &        \\
\ion{O}{i}   & 7773 & 8.83  &        &  8.88  & $+$0.05 &        &        &         &        &        &        \\
\ion{O}{i}   & 7774 & 8.83  &        &  8.82  & $-$0.01 &        &        &         &        &        &        \\
\ion{Eu}{ii} & 4129 & 0.51  &        &  0.47  & $-$0.04 &        &  0.46  & $-$0.05 &        &        &        \\
\ion{Eu}{ii} & 6645 & 0.51  &        &  0.56  & $+$0.05 &        &        &         &        &        &        \\
\noalign{\smallskip}
\hline
\end{tabular}
\label{tab:solarvalues}
\end{table*}
%------------------------------------------------------------------------------

%------------------------------------------------------------------------------
\subsubsection{Synthesis of the forbidden oxygen line at 6300\,{\AA}}

Synthesis of the forbidden oxygen line at 6300\,{\AA} has been done for the 
SOFIN spectra using the same methods and the same atomic data as in 
Bensby et al.~(\cite{bensby_syre}), apart from one thing: While we
in Bensby et al.~(\cite{bensby_syre}) used elliptical lineprofiles
to model the combined broadening of rotation and macroturbulence
we have here used radial-tangential (Rad-Tan) profiles instead.
The difference is small and is only reflected in a slight shift in the
absolute abundances. Since we always normalize our derived abundances to our 
own solar abundance this effect is of less importance
as is the choice of the standard abundances in col.~3 in 
Table~\ref{tab:solarvalues}.
                                     
We do, however, note that the solar oxygen abundance that we derive from
the forbidden [\ion{O}{i}] line at 6300\,{\AA} ($\rm \epsilon (O) = 8.71$ from
CES spectra and $\rm \epsilon (O) = 8.74$ from SOFIN spectra are in good
agreement with the new solar photospheric values by
Asplund et al.~(\cite{asplund_2004}). They derived
$\rm \epsilon (O) = 8.69$ using 3D models and $\rm \epsilon (O) = 8.73$
using the MARCS model. A strict comparison between our study and theirs
is, however, not straightforward since we have used a slightly \textit{lower}
$\log gf$-value for the [\ion{O}{i}] line and slightly \textit{higher}
$\log gf$-values for the blending \ion{Ni}{i} lines.
While we used $\log gf = -9.82$ for the [\ion{O}{i}] line they used
$\log gf = -9.72$. For the blending \ion{Ni}{i} lines we used the new
laboratory value, $\log gf = -2.11$, from Johansson et al.~(\cite{johansson})
which is split into $\log gf = -2.275$ for the $^{58}$Ni component and
$\log gf = -2.695$ for the $^{60}$Ni component
(see Bensby et al.~\cite{bensby_syre}). Asplund et al.~(\cite{asplund_2004})
used $\log gf = -2.31$ taken from their previous work
(Allende Prieto et al.~\cite{allendeprieto}).

%------------------------------------------------------------------------------
\subsubsection{Synthesis of the europium lines}

Determination of europium abundances have been done by synthesis of the
\ion{Eu}{ii} line at 6645\,{\AA} (in the FEROS spectra) and the
\ion{Eu}{ii} line at 4129\,{\AA} (in both FEROS and SOFIN spectra).
The synthesis was done in the same manner as for the [\ion{O}{i}] line
at 6300\,{\AA}. Europium has two isotopic components.  In the solar
system 47.8\,\% of Eu is in the form of $^{151}$Eu and 52.2\,\% in the
form of $^{153}$Eu.  Europium also shows large hyperfine splitting
which have to be taken into account in the abundance determination.
Linelists and $\log gf$-values for the hyperfine components of the Eu
lines have been kindly provided by C.~Sneden and are the same as those
used in the study by Lawler et al.~(\cite{lawler}). All atomic data
are given in Table~\ref{tab:atomdata}.

Examples of the synthesis of the two \ion{Eu}{ii} lines are shown in
Fig.~\ref{fig:eu_synt}, where the different isotopic hyperfine
components are indicated. Both Eu lines are also more or less blended
with other lines. These lines have been taken into account in the 
modelling of the spectra and are indicated in
Fig.~\ref{fig:eu_synt} as well.

The level of the continuum was in the case of the 4129\,{\AA} line
determined from the points just to the left of the \ion{Fe}{ii} line
at 4128.7\,{\AA} and just to the right of the \ion{Fe}{i} line at
4130.0\,{\AA}.  For the \ion{Eu}{ii} line at 6645 we used the
continuum points at 6644.7\,{\AA} and 6645.7\,{\AA} (see
Fig.~\ref{fig:eu_synt}).  For the instrumental broadening due to the
resolution of the FEROS and SOFIN instruments we adopted Gaussian
profiles with apropriate widths.  The combined broadening due to
macroturbulence and stellar rotation was determined by fitting
radial-tangential (Rad-Tan) profiles to the \ion{Fe}{ii} line at
4128.7\,{\AA} and the \ion{Ni}{i} line at 6643.6\,{\AA}, respectively.

%---------------------------------------------------------------------------
\begin{table*}
\caption{
        Estimates of the effects on the derived abundance ratios due to 
        internal (random) errors. The estimates are the average of four 
        stars (see Bensby et al.~\cite{bensby}). 
        }
\centering %\scriptsize
\begin{tabular}{lccccccc}       
\hline \hline\noalign{\smallskip}
                             &  
$\Delta\,T_{\rm eff}$        &  
$\Delta \log g$              &  
$\Delta\xi_{\rm t}$         & 
$\Delta$[Fe/H]               &
$\Delta\delta\gamma_{6}$     &
$\Delta W_{\lambda}/\sqrt{N}$ &
$\langle \sigma_{\rm rand} \rangle$ \\
                             &   
+70\,K                       &   
+0.1                         &
+0.15\,km\,s$^{-1}$          &
+0.1                         &
+50\,\%                      &
+5\,\%                       &
                             \\
\noalign{\smallskip}
\hline\noalign{\smallskip}
$\rm \Delta[\ion{Fe}{i}/H]$             &   $+0.05$ &   $-0.01$ &   $-0.03$ & $\pm0.00$ &  $\pm0.00$ &  $\pm0.00$ &  0.06   \\
$\rm \Delta[\ion{Fe}{ii}/H]$            &   $-0.02$ &   $+0.04$ &   $-0.03$ &   $+0.02$ &    $-0.03$ &  $\pm0.00$ &  0.06   \\
$\rm \Delta[\ion{Na}{i}/\ion{Fe}{i}]$   &   $-0.02$ &   $-0.01$ &   $+0.02$ & $\pm0.00$ &    $-0.01$ &    $+0.01$ &  0.03   \\
$\rm \Delta[\ion{Mg}{i}/\ion{Fe}{i}]$   &   $-0.02$ &   $-0.01$ &   $+0.02$ & $\pm0.00$ &    $-0.06$ &    $+0.01$ &  0.06   \\
$\rm \Delta[\ion{Al}{i}/\ion{Fe}{i}]$   &   $-0.03$ &   $-0.01$ &   $+0.02$ & $\pm0.00$ &    $-0.02$ &    $+0.01$ &  0.05   \\
$\rm \Delta[\ion{Si}{i}/\ion{Fe}{i}]$   &   $-0.04$ & $\pm0.00$ &   $+0.02$ & $\pm0.00$ &    $-0.04$ &  $\pm0.00$ &  0.05   \\
$\rm \Delta[\ion{Ca}{i}/\ion{Fe}{i}]$   & $\pm0.00$ &   $-0.02$ &   $+0.01$ & $\pm0.00$ &    $-0.01$ &  $\pm0.00$ &  0.03   \\
$\rm \Delta[\ion{Ti}{i}/\ion{Fe}{i}]$   &   $+0.03$ & $\pm0.00$ &   $+0.01$ & $\pm0.00$ &  $\pm0.00$ &  $\pm0.00$ &  0.03   \\
$\rm \Delta[\ion{Ti}{ii}/\ion{Fe}{i}]$  &   $-0.05$ &   $+0.04$ & $\pm0.00$ &   $+0.02$ &    $-0.03$ &  $\pm0.00$ &  0.07   \\
$\rm \Delta[\ion{Cr}{i}/\ion{Fe}{i}]$   & $\pm0.00$ & $\pm0.00$ &   $+0.02$ & $\pm0.00$ &  $\pm0.00$ &    $+0.01$ &  0.02   \\
$\rm \Delta[\ion{Cr}{ii}/\ion{Fe}{i}]$  &   $-0.06$ &   $+0.04$ & $\pm0.00$ &   $+0.02$ &    $-0.03$ &    $+0.01$ &  0.08   \\
$\rm \Delta[\ion{Ni}{i}/\ion{Fe}{i}]$   & $\pm0.00$ & $\pm0.00$ &   $+0.01$ & $\pm0.00$ &  $\pm0.00$ &  $\pm0.00$ &  0.02   \\
$\rm \Delta[\ion{Zn}{i}/\ion{Fe}{i}]$   &   $-0.03$ & $\pm0.00$ &   $-0.01$ &   $+0.02$ &    $-0.05$ &    $+0.02$ &  0.06   \\
$\rm \Delta[\ion{Y}{ii}/\ion{Fe}{i}]$   &   $-0.06$ &   $+0.03$ &   $-0.01$ &   $+0.02$ &    $-0.03$ &    $+0.01$ &  0.08   \\
$\rm \Delta[\ion{Ba}{ii}/\ion{Fe}{i}]$  &   $-0.03$ &   $+0.01$ &   $-0.02$ &   $+0.04$ &  $\pm0.00$ &    $+0.01$ &  0.06   \\
\noalign{\smallskip}
\hline
\end{tabular}
\label{tab:errors}
\end{table*}
%------------------------------------------------------------------------------

%==============================================================================
\subsection{Random errors}

The effects on the derived abundances due to random (internal) errors
were estimated in Bensby et al.~(\cite{bensby}). By changing $T_{\rm
eff}$ by +70\,K, $\log g$ by +0.1, $\xi_{\rm t}$ by
+0.15\,km\,s$^{-1}$, [Fe/H] by +0.1, and the correction term to the
Uns\"old approximation of the Van der Waals damping by +50\,\%, the
effects were studied on four stars. The average of the total random
error from these four stars are listed in Table~\ref{tab:errors} for
various abundance ratios.  It should be noted that these estimates are
made under the assumptions that the different error sources are
uncorrelated, which might not be completely true. Errors in the
effective temperature will for instance also show up in the plot of
\ion{Fe}{i} abundances versus line strength, i.e. in the tuning of the
microturbulence.  Hence, if this erroneous $T_{\rm eff}$ was used in
an analysis the researcher would adjust $\xi_{\rm t}$ to achieve
equilibrium and in this way probably partly compensate the erroneous
$T_{\rm eff}$ with an erroneous $\xi_{\rm t}$.  However, the
equilibrium would not be as good as if the correct parameters had been
used.  Therefore the total errors in Table~\ref{tab:errors} should be
treated as maximum internal errors.  That these really are maximum
errors is further reflected in the tight abundance trends we obtain
where the scatter within each stellar population are lower than these
estimated errors.

%==============================================================================
\subsection{Systematic errors: Comparison to other studies}
\label{sec:errors}

Systematic errors are more difficult to examine.  In Bensby et
al.~(\cite{bensby}) we compared our solar equivalent widths (measured
in a FEROS spectrum) to those in Edvardsson et al.~(\cite{edvardsson})
and found good agreement.  The comparisons between our SOFIN and UVES
equivalent widths to our FEROS equivalent widths (see
Fig.~\ref{fig:ew_sof_fer}) showed that there are small offsets
present. Since we normalize the abundances from the different
spectrographs separately (see Sect.~\ref{sec:solaranalysis}) we do not
expect there to be any offsets in the derived abundances between the
different data sets.

In Bensby et al.~(\cite{bensby}) we compared
our derived abundances for a few stars to abundances from other works and
found, generally, good agreement. 
In the present sample (i.e., all 102 stars) we have four stars in
common with Reddy et al.~(\cite{reddy}); HIP~11309 (HD~15029),
HIP~85007 (HD~157466), HIP~92270 (HD~174160), and HIP~118010
(HD~224233).  In Fig.~\ref{fig:reddy_comp} we compare our abundances
with theirs. For three of the stars the differences in [X/Fe], X being
any of the elements considered, are small apart from for one or two of
the elements. For HIP~11309 the mean difference is
$+0.005\pm0.048$\,dex, for HIP~85007 the mean difference is
$+0.019\pm0.063$\,dex, and for HIP~92270 the difference is
$-0.016\pm0.058$\,dex. We note that for these three stars it is
essentially two elements that contribute to the scatter, Y and Ba.
These two elements also show systematic differences between the two
studies in that we always derive larger Ba abundances and smaller Y
abundances than Reddy et al.~(\cite{reddy}).  If Ba and Y elements are
removed from the calculation the resulting mean differences and
scatters become $+0.005\pm0.039$, $+0.020\pm0.036$, $-0.024\pm0.045$,
for HIP~11309, HIP~85007, and HIP~92270, respectively.  For
HIP~118010, however, the scatter around the mean difference is larger,
0.091\,dex, and appear to be more random in nature. The scatter is not
decreased when Ba and Y are removed. The reason for this difference
lays in that we use an effective temperature that is 200\,K lower than
the one Reddy et al.~(\cite{reddy}) use. We have adopted a $T_{\rm
eff}$ of 5795\,K and Reddy et al.~(\cite{reddy}) use 5609\,K.  Both
studies use the same surface gravity, 4.17\,dex. This combination
results in that we derive $\rm [Fe/H]=-0.07$ and they arrive at
$-0.24$\,dex.  If we use Table~\ref{tab:errors} to estimate the
correction for this difference we arrive at abundances that are very
similar to those of Reddy et al.~(\cite{reddy}).  For the other three
stars the stellar parameters are identical, within the errors, between
the studies.  From this comparison we conclude that our data and Reddy
et al.~(\cite{reddy}) are in good agreement and when differences occur
they can be understood and corrected for.  This means that it is
possible to combine the results from our studies with those of Reddy
et al.~(\cite{reddy}) when there is a need for large data samples,
e.g. when comparing models of chemical evolution to data.
%------------------------------------------------------------------------------
\begin{figure}
 \resizebox{\hsize}{!}{\includegraphics{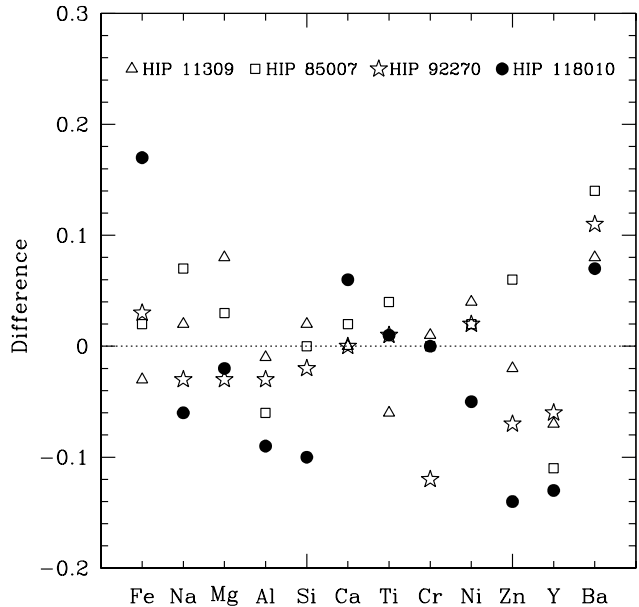}}
\caption{
        Comparison of our abundances with Reddy et al.~(\cite{reddy}) for
        four stars in common. The differences for each individual star
        are marked by symbols as indicated in the figure. 
        The differences are defined as
        $\rm [X/Fe]_{This\,\,Work} - [X/Fe]_{Reddy}$ except for Fe which
        is $\rm [Fe/H]_{This\,\,Work} - [Fe/H]_{Reddy}$.
        Note that for four
        elements, Ca, Ti, Cr, Ni, the differences between two stars are
        identical and hence we only see three symbols.
         }
\label{fig:reddy_comp}
\end{figure}
%------------------------------------------------------------------------------

Our stellar sample have eight stars in common with the studies by
Mashonkina \& Gehren~(\cite{mashonkina1}, \cite{mashonkina}) and
Mashonkina et al.~(\cite{mashonkina3}).  In
Fig.~\ref{fig:mashonkina_comp} we show a comparison between our Fe,
Ba, and Eu abundances with theirs for these eight stars.  Except for
two Ba abundances (HIP~699 and HIP~107975) the differences are small.
For [Fe/H] the difference is $0.00\pm0.04$\,dex, for [Ba/Fe] the
difference is $0.10\pm0.10$\,dex, and for [Eu/Fe] the difference is
$0.01\pm0.04$\,dex.  Excluding HIP~699 and HIP~107975 the difference
for [Ba/Fe] shrinks to $0.05\pm0.02$\,dex.  The reason for the
deviating Ba abundances for these two stars is difficult to resolve.
If we look at the stellar parameters we see that our effective
temperatures and surface gravities are higher than what Mashonkina and
collaborators have used.  For HIP~699 we have $T_{\rm eff} = 6250$\,K and
$\log g=4.19$ while they have $T_{\rm eff} = 6150$\,K and $\log g=4.06$,
and for HIP~107975 we have $T_{\rm eff} = 6460$\,K and $\log g=4.06$
while they have $T_{\rm eff} = 6310$\,K and $\log g=3.94$.  From
Table~\ref{tab:errors} we see that this does not resolve the
discrepancies for these stars. If we were to lower our effective
temperatures our [Ba/Fe] ratios would actually increase and make the
differences larger. A lowering of the surface gravities would on the
other hand lower our [Ba/Fe] ratios but not sufficiently.  A lowering
of $\log g$ by 0.1\,dex would only result in a 0.01\,dex lowering of
[Ba/Fe].  It is worth noting that none of these stars have abundances
that make them deviate from the abundance trends that the rest
of our stars outline (see Sect.~\ref{sec:outliers} for stars that do).
The otherwise good agreement to the works by Mashonkina and collaborators
should justify combinations of stars and elemental abundances from their and
our works when needing larger data sets. From the good agreement between
our and Mashonkina and collaborators' Eu abundances we estimate
systematic errors to be small for Eu, around or below 0.05\,dex in both [Eu/Fe]
and [Eu/H].

%==============================================================================
\section{Stellar ages}
\label{sec:ages}

%------------------------------------------------------------------------------
\begin{figure}
 \resizebox{\hsize}{!}{
            \includegraphics[bb=18 144 592 555,clip]{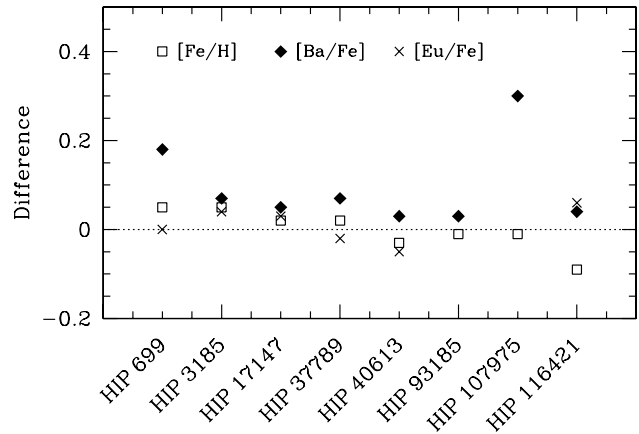}}
\caption{
        Comparison of abundances for eight stars that we
        have in common with the studies by
        Mashonkina \& Gehren~(\cite{mashonkina1}, \cite{mashonkina})
        and Mashonkina et al.~(\cite{mashonkina3}).
        The differences are defined as our abundances minus theirs.
        Note that the Mashonkina studies have no Eu abundances
        for two of the stars: HIP~93185 and HIP~107975.
         }
\label{fig:mashonkina_comp}
\end{figure}
%------------------------------------------------------------------------------

In Bensby et al.~(\cite{bensby}) we used the Salasnich et
al.~(\cite{salasnich}) and Girardi et al.~(\cite{girardi}) isochrones
to estimate the stellar ages. The Yoshii-Yale (Y$^2$) isochrones (Kim
et al~\cite{kim}; Yi et al.~\cite{yi}) are distributed together with
interpolation routines that makes it possible to construct a set of
isochrones with any metallicity and $\alpha$-enhancement.  Other
isochrone sets are tabulated for fixed values of these
parameters. These fixed values may not necessarily coincide with the
analyzed data. Since $\alpha$-enhancement varies for our stars we have
opted to use the Y$^2$-isochrones.  At sub-solar [Fe/H] we used
different $\alpha$-enhancements for thin and thick disk stars
according to our observations (see Table~\ref{tab:isochrones}).  The
most likely age was then estimated for each star from isochrones
plotted in the $T_{\rm eff}$\,--\,M$_{\rm V}$ plane, using Hipparcos
parallaxes and our spectroscopic temperatures.  Lower and upper age
limits were estimated from the plots as well by taking the errors in
the parallaxes and effective temperaures into account.  Stellar ages
were determined for the new stars in this study and, in order to get a
consistent age determination for the whole stellar sample, those in Bensby
et al.~(\cite{bensby}) as well.  The ages and their lower and upper
limits are given in Table~\ref{tab:ages}. Other methods
to derive stellar ages from isochrones exist.  However, our age
estimates are virtually identical to those obtained with more
sophisticated methods.  Our method, on the other hand, probably
underestimates the uncertainties of the derived ages 
(Rosenkilde J\o rgensen, private comm.).

%------------------------------------------------------------------------------
\begin{table}[t]
\centering
\caption{
        Metallicities and $\alpha$-enhancements for the Y$^2$ isochrones
        that were used in the age determination.
        }
\centering %\scriptsize
\begin{tabular}{rrr|rrr}
\hline \hline\noalign{\smallskip}
[Fe/H]  & \multicolumn{2}{c|}{[$\alpha$/Fe]} 
        & [Fe/H]  
        & \multicolumn{2}{c}{[$\alpha$/Fe]}  \\
        & Thin  
        & Thick             
        &         
        & Thin  
        & Thick            \\
\noalign{\smallskip}
\hline\noalign{\smallskip}
\noalign{\smallskip}
$-$0.90  &           &  +0.30  & $-$0.20  & +0.05     &  +0.15  \\
$-$0.80  &           &  +0.30  & $-$0.10  & +0.03     &  +0.10  \\
$-$0.70  & +0.15     &  +0.30  &    0.00  & +0.03     &  +0.03  \\
$-$0.60  & +0.13     &  +0.30  &   +0.10  & +0.03     &  +0.03  \\
$-$0.50  & +0.12     &  +0.30  &   +0.20  & +0.03     &  +0.03  \\
$-$0.40  & +0.10     &  +0.30  &   +0.30  & +0.03     &  +0.03  \\
$-$0.30  & +0.07     &  +0.20  &   +0.40  & +0.03     &  +0.03  \\
 \noalign{\smallskip}
\hline
\end{tabular}
\label{tab:isochrones}
\end{table}
%------------------------------------------------------------------------------
%------------------------------------------------------------------------------
\begin{table}[t]
\centering
\caption{
        Age estimates for the stars in this study and those in
        Bensby et al.~(\cite{bensby}). The minimum (Min age) and maximum
        (Max age)
        ages are based on the uncertainties in the Hipparcos
        parallaxes and $\pm 70$\,K in the effective temperatures.
        The full table is available via anonymous ftp at
        \texttt{cdsarc.u-strasbg.fr} (130.79.128.5) or via
        \texttt{cdsweb.u-strasbg.fr/cgi-bin/qcat?J/A+A/}
        }
\centering %\scriptsize
\begin{tabular}{ccccc}
\hline \hline\noalign{\smallskip}
HIP     &
Mem.    &
Age     &
Min age &
Max age \\
\noalign{\smallskip}
\hline\noalign{\smallskip}
\noalign{\smallskip}
  699   &  1  &  3.8  &  3.4  &  4.4  \\
  910   &  1  &  5.5  &  5.0  &  5.8  \\
           $\vdots$
        &  $\vdots$
        &  $\vdots$
        &  $\vdots$
        &  $\vdots$ \\
\noalign{\smallskip}
\hline
\end{tabular}
\label{tab:ages}
\end{table}
%------------------------------------------------------------------------------

The mean ages of the thin and thick disk samples (including the new age
determinations for the stars from Bensby et al.~\cite{bensby})
are $4.3\pm2.6$\,Gyr and $9.7\pm3.1$\,Gyr, respectively. The mean age
for the four stars with intermediate kinematics is $6.7\pm2.0$\,Gyr.

Figure~\ref{fig:agemet} shows [Fe/H] as a function of age for all 102 stars.
For the thick disk stars there might exist a relation between age and
metallicity. Thick disk stars with [Fe/H]\,$\leq -0.4$ have a mean age of
$11.6\pm2.9$\,Gyr, those with $\rm -0.4<[Fe/H]\leq-0.2$ have a mean age
of $8.1\pm2.7$\,Gyr, and those with [Fe/H]\,$>-0.2$ have a mean age of
$7.7\pm1.4$\,Gyr.
This indicates that star formation could have continued in the thick disk
for quite some time, up to about 2--3\,Gyr. This conclusion is however 
uncertain due to the large spread in the stellar ages and especially 
to the rather small stellar
sample we have here. The potential trend between age and metallicity in the
thick disk is, however, very similar to the results we find in a study
of ages and metallicities (based on Str\"omgren $uvby$ photometry) of a
larger sample of thick disk stars (Bensby et al.~\cite{bensby_amr}).
In that study we found a possible age-metallicity relation in the thick disk
and also that it might have taken 2--3\,Gyr for the thick disk to reach
[Fe/H]\,$\approx -0.4$ and a further 2--3\,Gyr to reach solar metallicites.

%------------------------------------------------------------------------------
\begin{figure}[t]
\resizebox{\hsize}{!}{
        \includegraphics{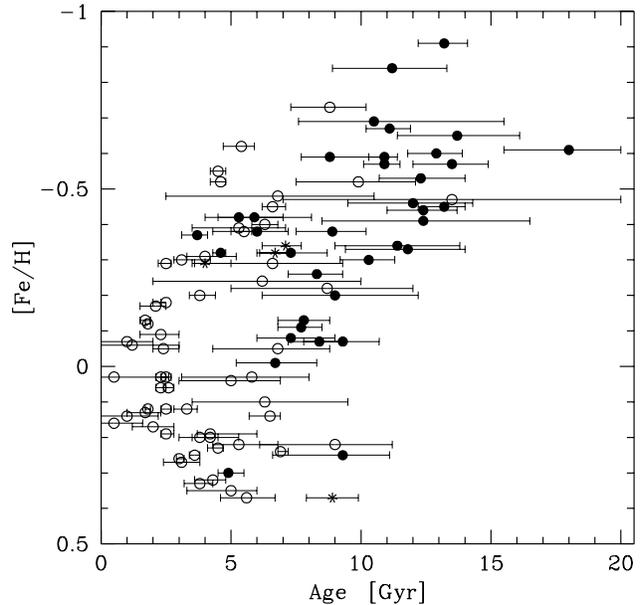}}
\caption{
        [Fe/H] versus age for the 36 stars from this study
        and the 66 stars from Bensby et al.~(\cite{bensby}). Thin disk and
        thick disk stars are
        marked by open and filled circles, respectively. Transition objects
        are marked by asterisks.
         }
\label{fig:agemet}
\end{figure}
%------------------------------------------------------------------------------

%------------------------------------------------------------------------------
\begin{table}[t]
\centering
\caption{
        Derived abundances relative to hydrogen. 
        Each element has three columns, 
        mean abundance ([X/H]), standard deviation of the mean abundance
        ($\sigma_{\rm [X/H]}$), and the number of spectral lines ($N_{\rm X}$) 
        that were used in the abundance analysis (Cols. 4 and onwards). 
        The abundances have been normalized with respect to the Sun  
        (see Sect.~\ref{sec:solaranalysis} and Table~\ref{tab:solarvalues}). 
        Column~1 gives the Hipparcos number;
        col.~2 indicates with which spectrograph the star was observed
        (S=SOFIN, F=FEROS, U=UVES);
        col.~3 indicates if the star
        belongs to the thin disk (Mem.=1), the thick disk (Mem.=2), or a
        ``transition object" (Mem.=3). The full
        table is available in electronic form at the CDS via anonymous ftp to
        \texttt{cdsarc.u-strasbg.fr} (130.79.128.5) or via
        \texttt{cdsweb.u-strasbg.fr/cgi-bin/qcat?J/A+A/}
        }
\centering %\scriptsize
\begin{tabular}{ccccccc}
\hline \hline\noalign{\smallskip} HIP & Inst.
        & Mem.  
        & [\ion{Fe}{i}/H] 
        & $\sigma_{\rm [\ion{Fe}{i}/H]}$ 
        & $N_{\rm \ion{Fe}{i}}$ 
        & $\ldots$ \\
\noalign{\smallskip} \hline\noalign{\smallskip} \noalign{\smallskip}
  699 & S & 1 & $-$0.20 & 0.06 & 79 & $\ldots$ \\ 
  910 & S & 1 & $-$0.36 & 0.07 & 68 & $\ldots$ \\
 $\vdots$ & $\vdots$ & $\vdots$ & $\vdots$ & $\vdots$ & $\vdots$ & $\ddots$ \\
\noalign{\smallskip} \hline
\end{tabular}
\label{tab:abundances}
\end{table}
%------------------------------------------------------------------------------

We note that at the highest metallicities ([Fe/H]\,$>0.2$) there is a lack
of young stars (ages lower than 3\,Gyr). Given that there is ongoing star
formation in the metal-rich thin disk today this is clearly not a 
representative picture. This apparent trend is probably due to selection 
effects in our 
sample, i.e. only contructed from F and G dwarf stars and not including
earlier type stars, and not a feature of the Galactic thin disk.

%==============================================================================
\section{Abundance results} \label{sec:results}

\subsection{Strengthening the $\alpha$- and iron-peak 
        element trends} 

In Fig.~\ref{fig:x_fe} we show the abundance trends relative to Fe for
all 102 stars. The new stars from the northern sample confirm and
extend the trends that we presented in Bensby et al.~(\cite{bensby},
\cite{bensby_syre}).  No major novelties are found so these trends
will only be briefly described and the reader is directed to Bensby et
al.~(\cite{bensby},~\cite{bensby_syre}) for further discussions and
comparisons to other works.

%------------------------------------------------------------------------------
\begin{figure*}[ht]
\resizebox{\hsize}{!}{
        \includegraphics[bb=18 200 592 475,clip]{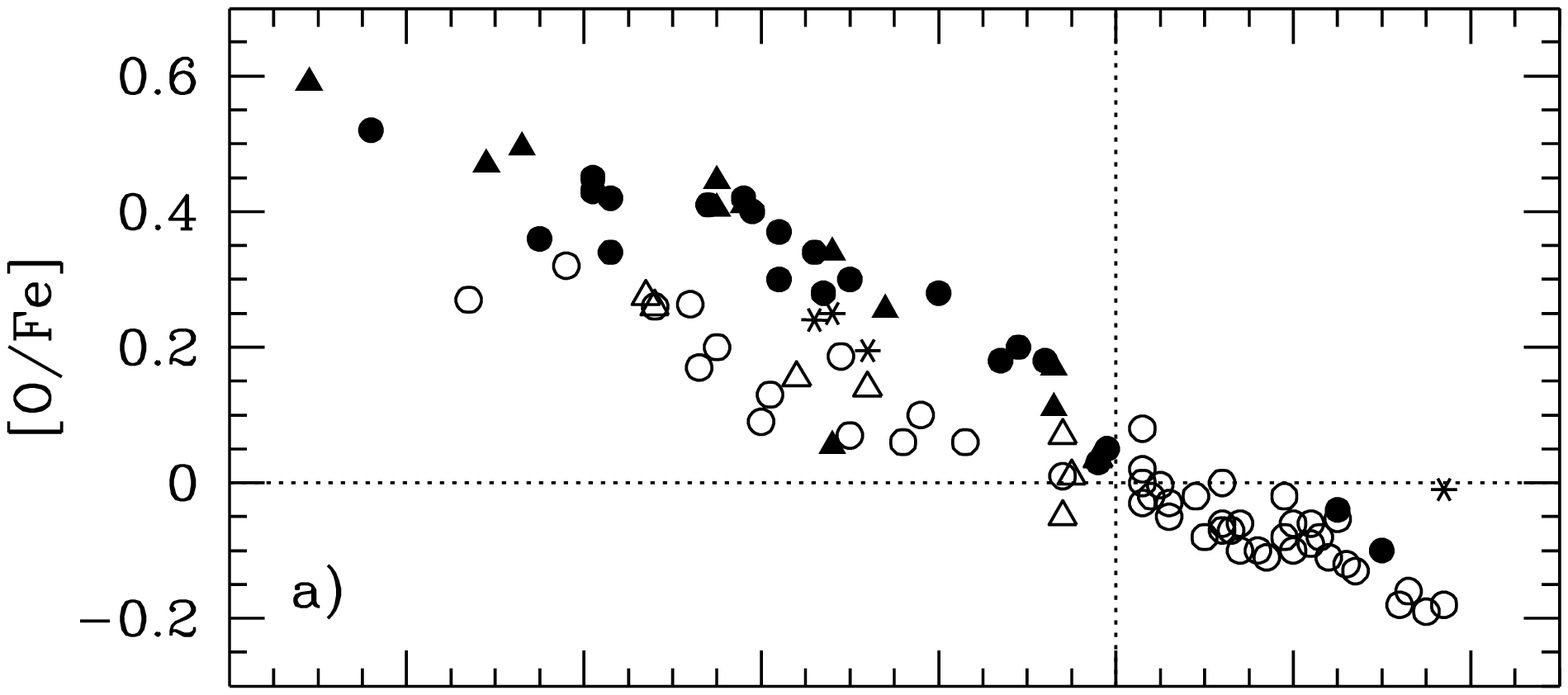} \hspace{10mm}
        \includegraphics[bb=18 200 592 475,clip]{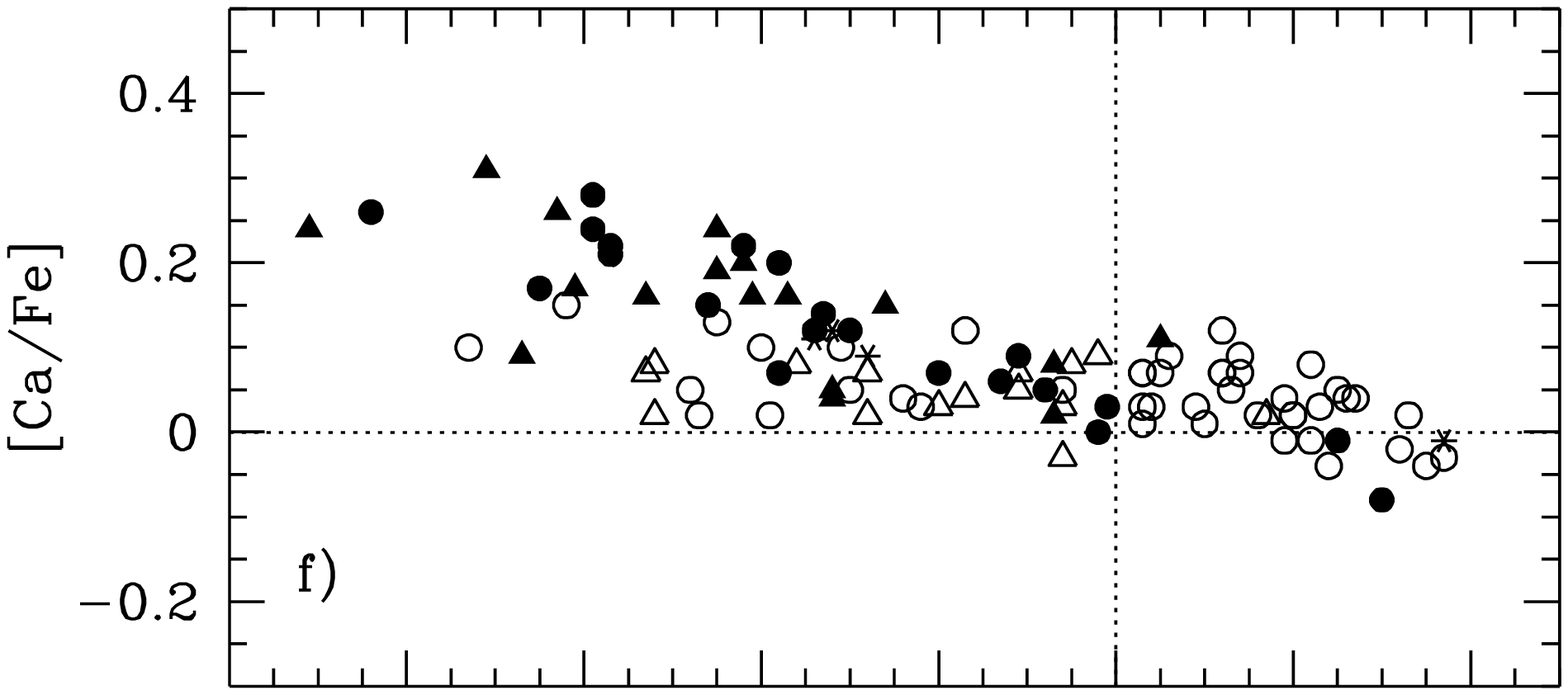}}

\resizebox{\hsize}{!}{
        \includegraphics[bb=18 200 592 450,clip]{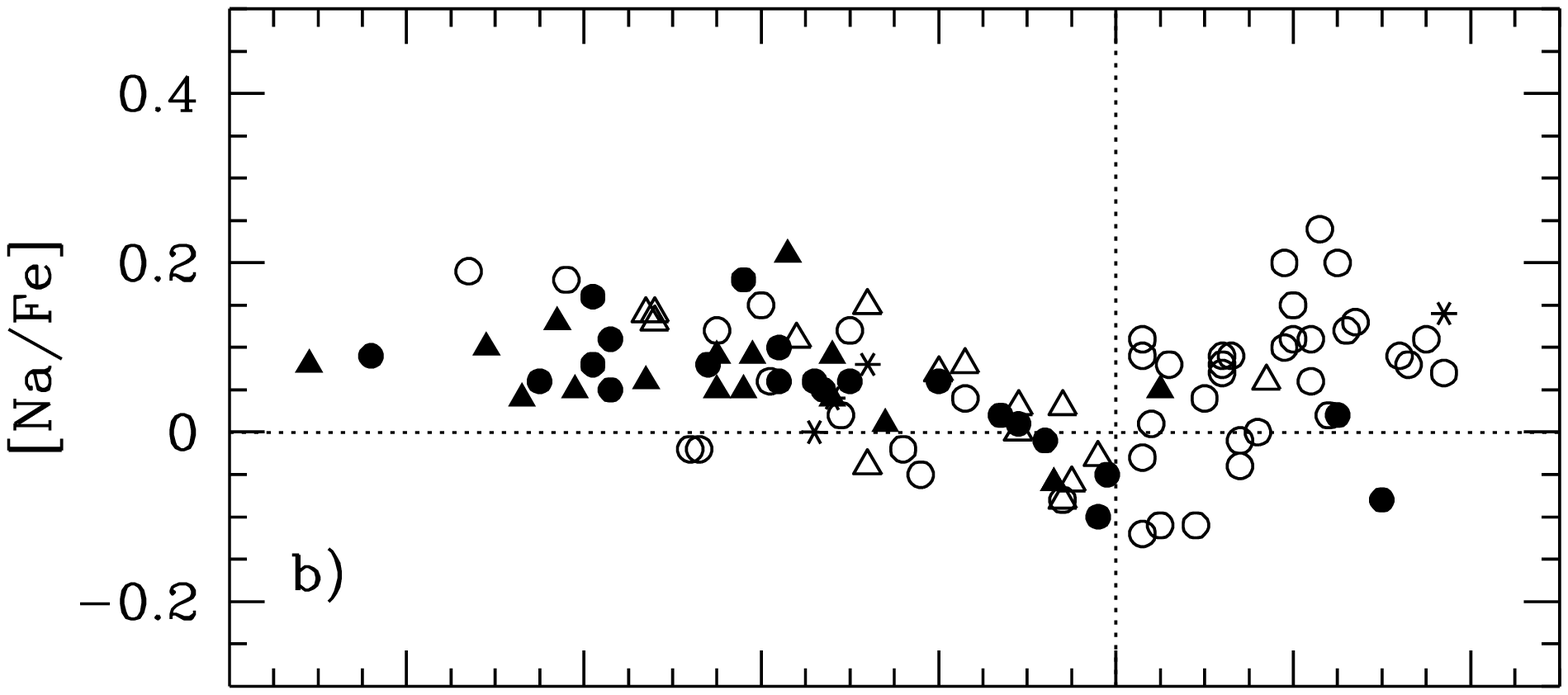} \hspace{10mm}
        \includegraphics[bb=18 200 592 450,clip]{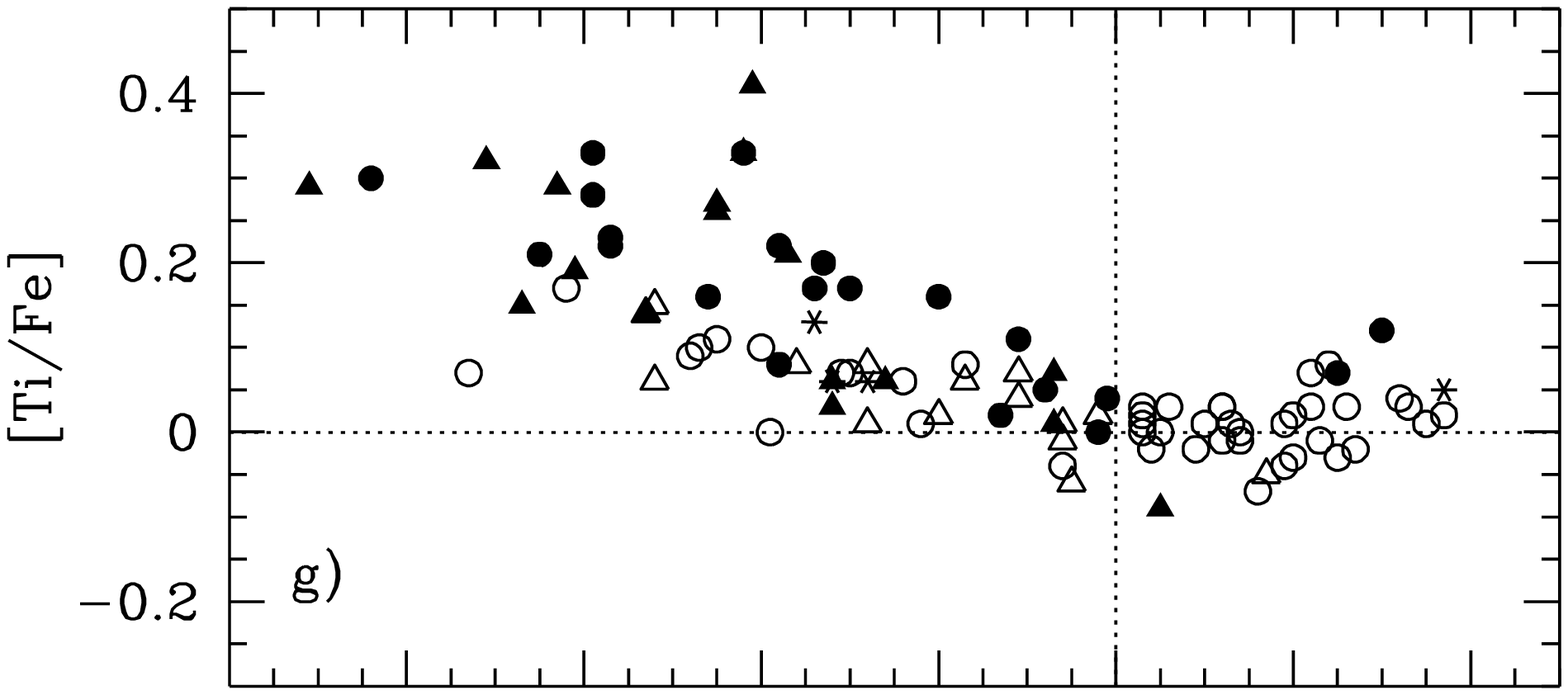}}

\resizebox{\hsize}{!}{
        \includegraphics[bb=18 200 592 450,clip]{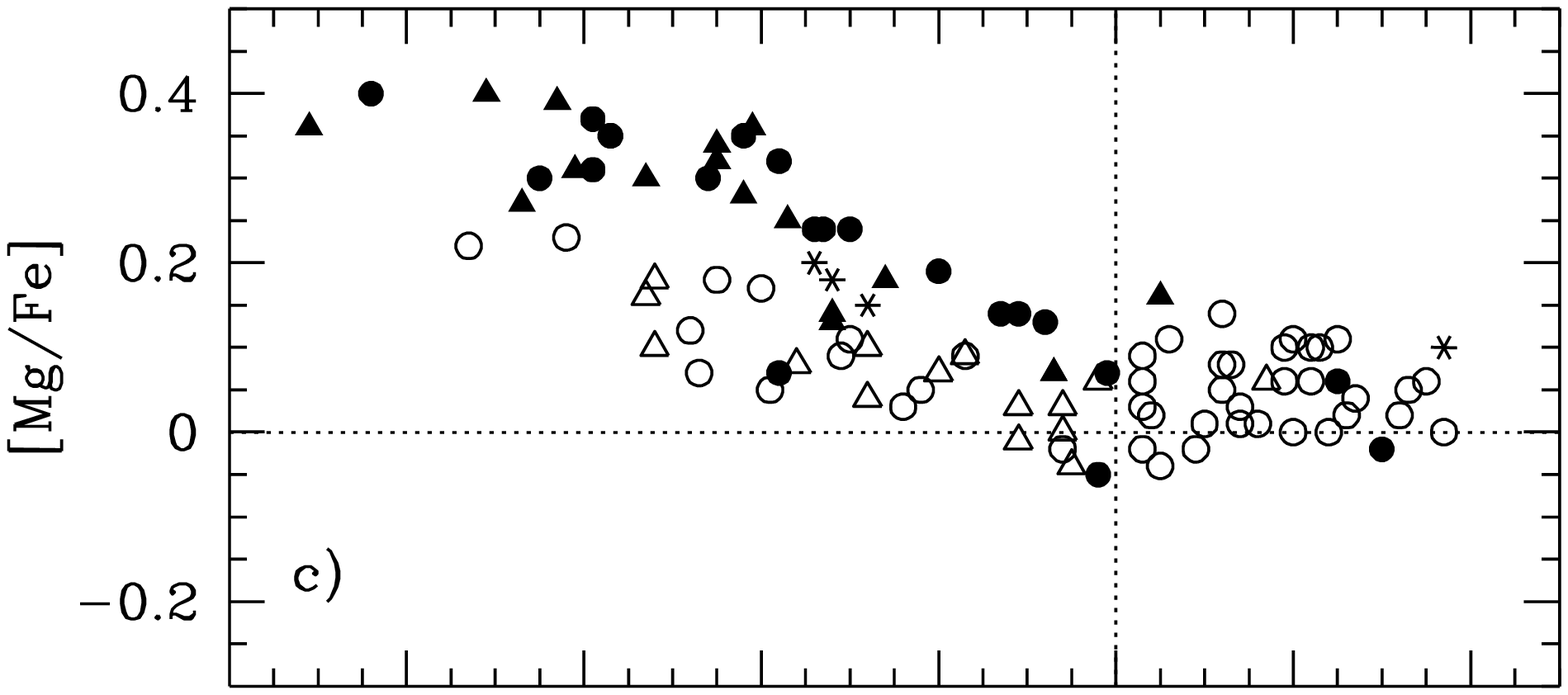} \hspace{10mm}
        \includegraphics[bb=18 200 592 450,clip]{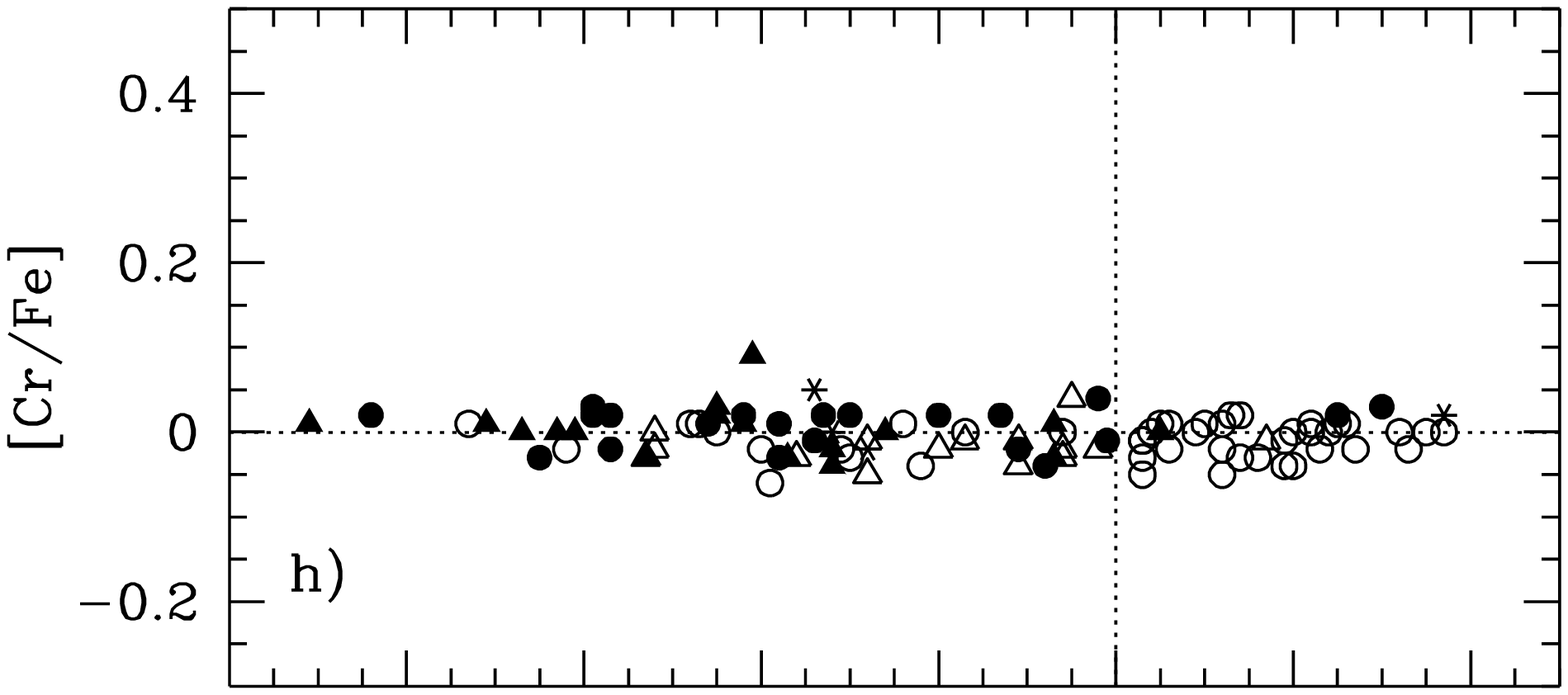}}

\resizebox{\hsize}{!}{
        \includegraphics[bb=18 200 592 450,clip]{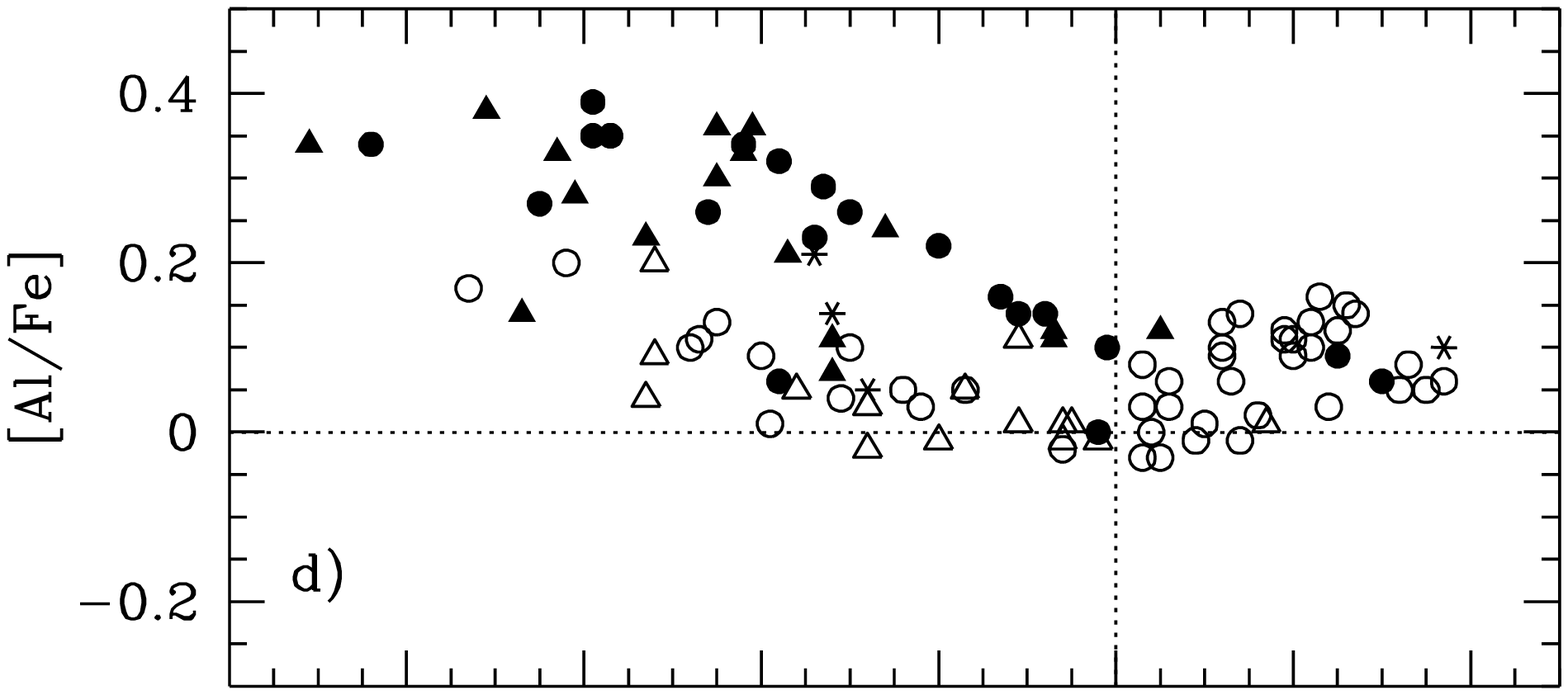} \hspace{10mm}
        \includegraphics[bb=18 200 592 450,clip]{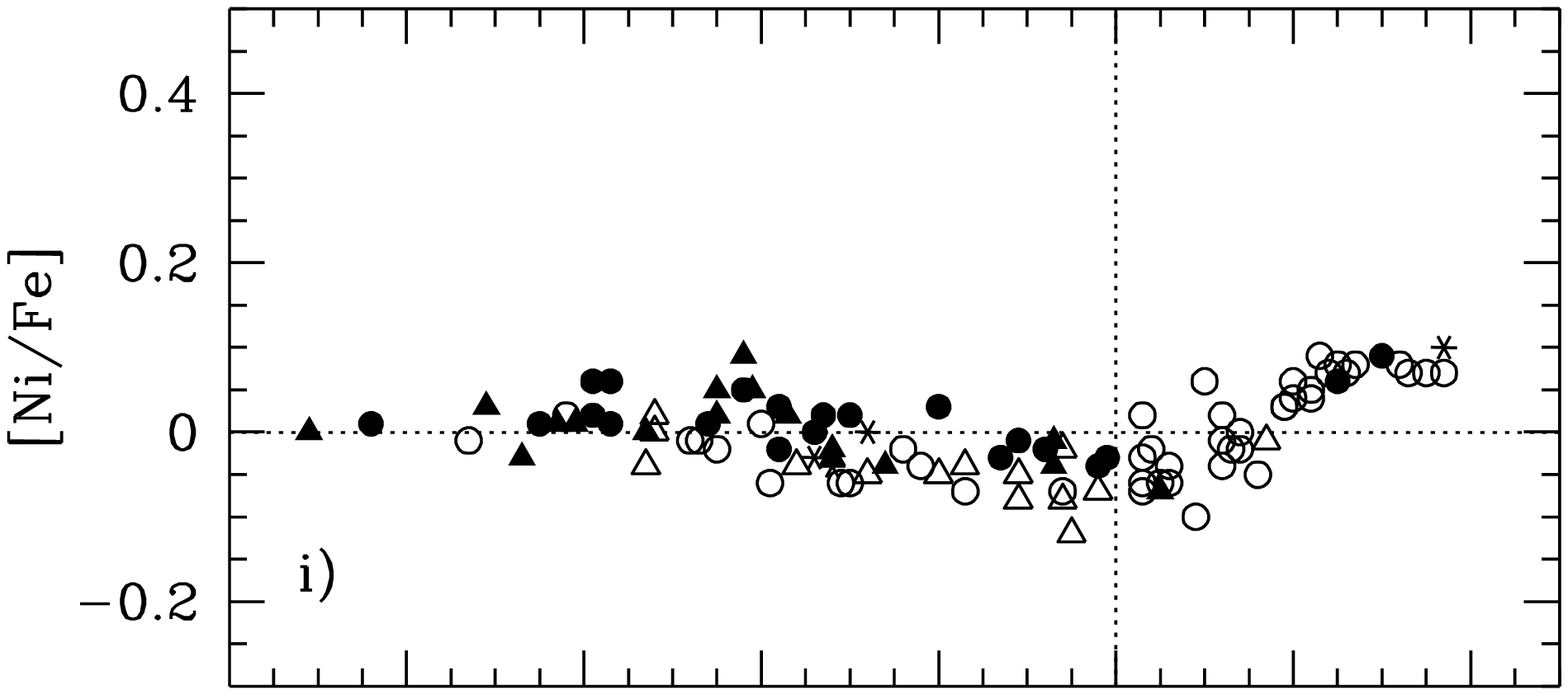}}

\resizebox{\hsize}{!}{
        \includegraphics[bb=18 144 592 450,clip]{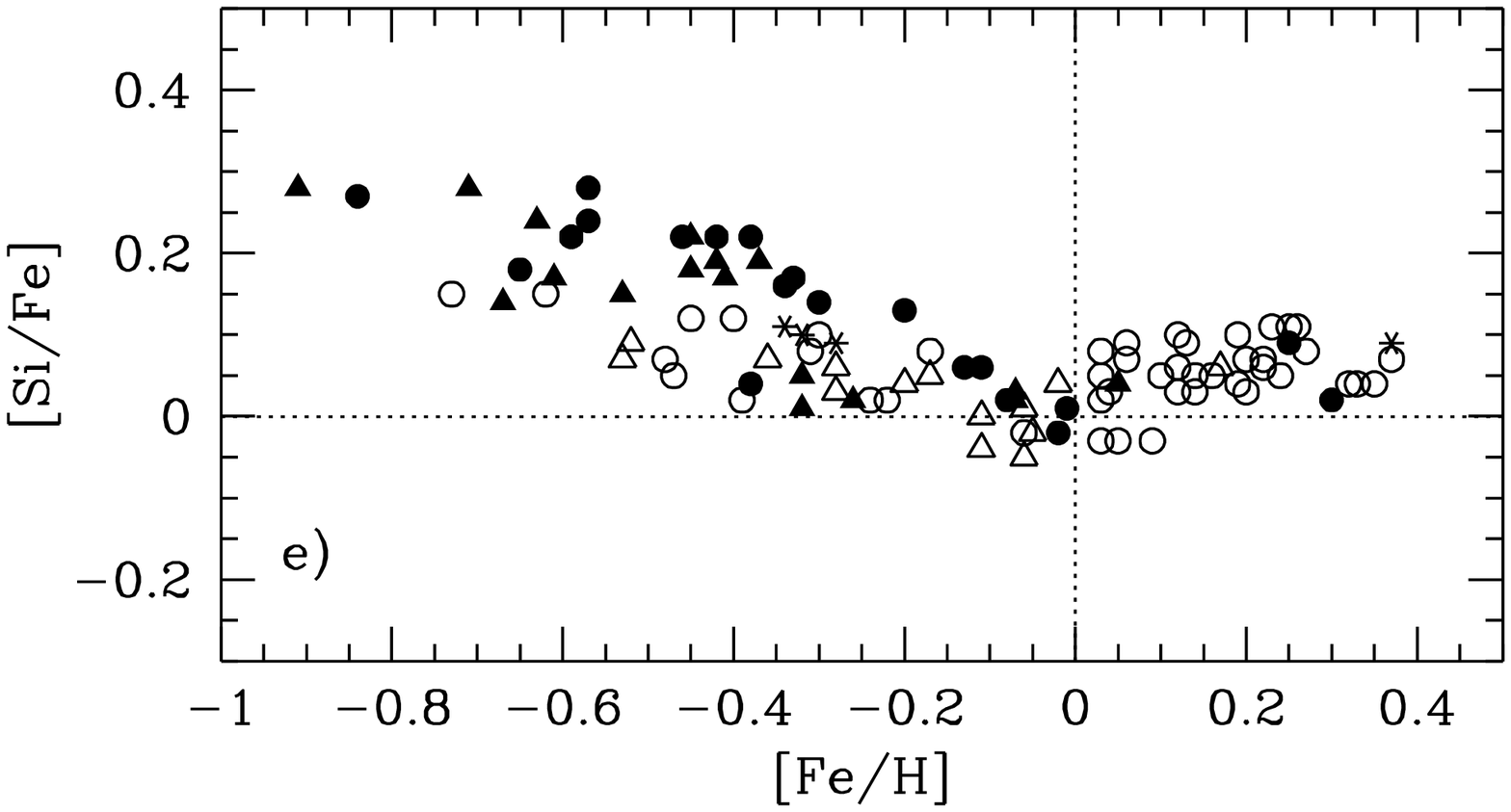} \hspace{10mm}
        \includegraphics[bb=18 144 592 450,clip]{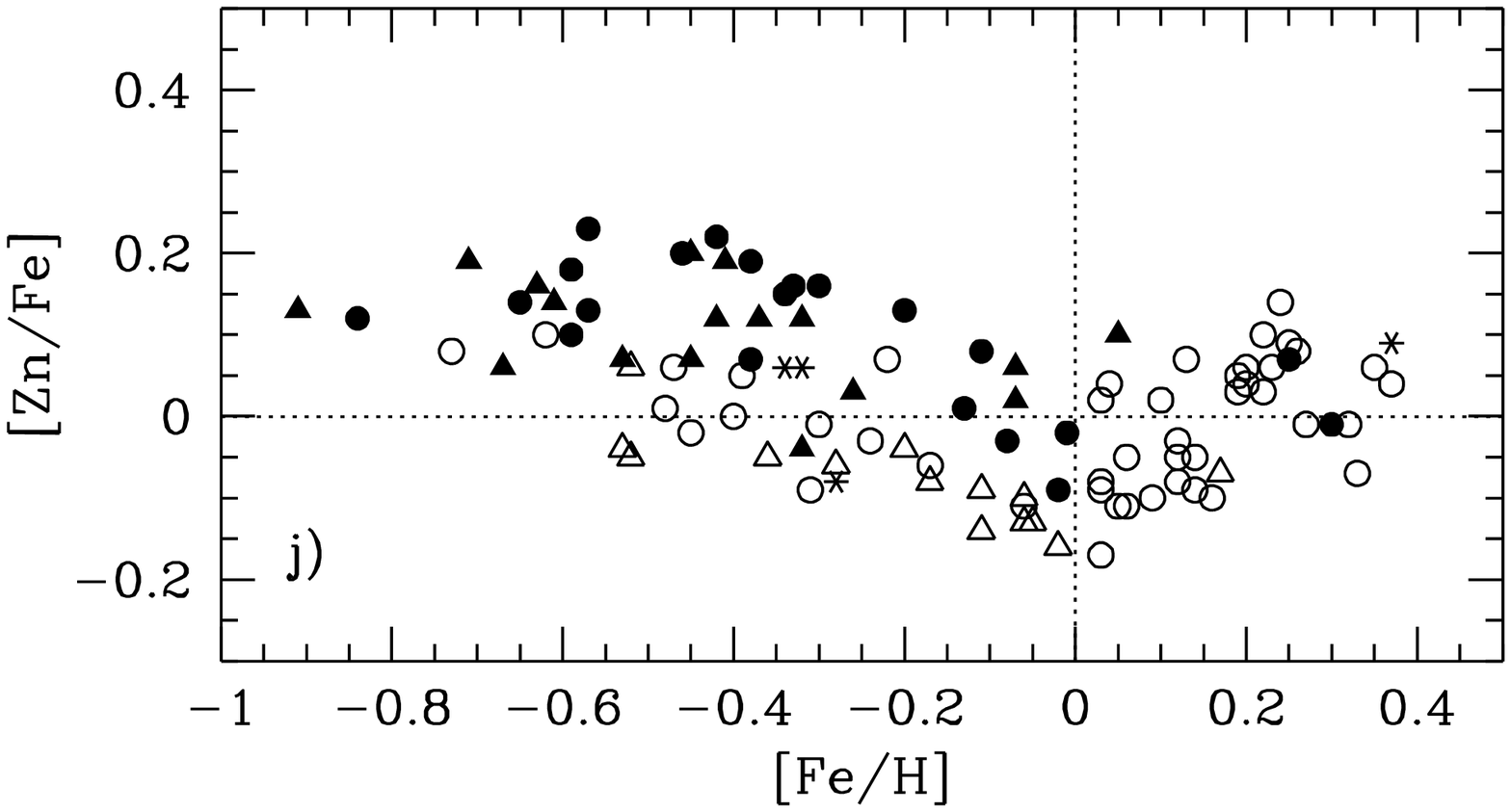}}

\caption{
        Elemental abundances relative to Fe. Dotted lines indicate
        solar values. Thin disk and thick disk stars are marked by 
        empty and filled circles, respectively. Stars from 
        Bensby et al.~(\cite{bensby}, \cite{bensby_syre}) are marked 
        by circles and stars from the new northern sample by triangles.
        Transition objects are marked by asterisks.
         }
\label{fig:x_fe}
\end{figure*}
%------------------------------------------------------------------------------

\paragraph{O, Mg, Si, Ca, and Ti:}
Thin and thick disk stars are clearly separated and show tight and
distinct trends. The down-turns (or ``knees") in the [O/Fe] and
[Mg/Fe] trends for the thick disk stars are especially prominent (see
Fig.~\ref{fig:x_fe}a and c).  There is no doubt that the locations of
these ``down-turns" are at $\rm [Fe/H] \approx -0.4$ after which
[O/Fe] and [Mg/Fe] decline toward solar values.  At lower [Fe/H] the
trends are essentially flat showing constant values of [Mg/Fe]\,$\sim
+0.35$ and [O/Fe]\,$\sim +0.4$.

Although not as prominent as for O and Mg, these features are also
clearly present for the other three $\alpha$-elements; Si, Ca, and Ti.

\paragraph{Na and Al:}
The appearance of the thin and thick disk [Na/Fe] trends
(Fig.~\ref{fig:x_fe}b) show an inverted behaviour compared to the
$\alpha$-elements. Instead, the stars in thin disk seem to be more
abundant in Na than those in the thick disk.  At first glance this
[Na/Fe] trend appear to be different from that in Bensby et
al.~(\cite{bensby}) where we found that [Na/Fe] trends for 
the thin and thick disk
 appear to be merged. The tighter [Na/Fe] trends in
this study (more stars are now used to trace the trends) indicate that also
the Na abundances are distinct between the two disk populations, even if 
not as well separated as for the $\alpha$-elements.

The Al trends show the same type of trends as the $\alpha$-elements
(see Fig.~\ref{fig:x_fe}d). This supports our findings in 
Bensby et al.~(\cite{bensby}) that Al and the $\alpha$-elements 
are produced in the same environments and have been dispersed into 
the interstellar medium on the same time-scales, i.e., 
the SN\,II events. Also, McWilliam~(\cite{mcwilliam}) noted that
Al, from a phenomological point of view, can be classified as an
$\alpha$-element. 

\paragraph{Cr and Ni:}
Cr varies tightly in lock-step with Fe (see Fig.~\ref{fig:x_fe}h).  No
trends are seen and the thin and thick disk stars are well mixed which
strongly emphasize the common origin for Cr and Fe.

In Bensby et al.~(\cite{bensby}) we found that the Ni and Fe below
solar metallicities evolve roughly in lockstep
(i.e., [Ni/Fe]\,$\approx$\,0).  At [Fe/H]\,$\sim$\,0 the [Ni/Fe] trend
then showed a prominent up-turn that had not been seen in previous
studies. This deviation from a flat [Ni/Fe] trend has an impact on the
oxygen abundances that are derived from the forbidden [\ion{O}{i}]
line at 6300\,{\AA} since this line is heavily blended by two
\ion{Ni}{i} lines (see Bensby et al.~\cite{bensby_syre}). In the new
sample there are only two stars with [Fe/H]\,$>0$. They do, however,
follow the same trend as found in Bensby et al.~(\cite{bensby}) and
show an increased [Ni/Fe] (see Fig.~\ref{fig:x_fe}i). The now larger
number of stars at [Fe/H]\,$<0$ also indicates that it is possible
that the [Ni/Fe] trend at these metallicities actually is not flat. We
see a slight overall decrease in [Ni/Fe] when going to higher [Fe/H],
and at [Fe/H]\,$=0$ there is an underabundance of Ni relative Fe of
about 0.05\,dex. There is also a weak tendency that the thick disk
stars are more abundant in Ni than the thin disk stars.

\paragraph{Zn:}
The [Zn/Fe] trend is tight and in accordance with Bensby et
al.~(\cite{bensby}) (see Fig.~\ref{fig:x_fe}j). This is somewhat
surprising since the trend for the new stars is based on one
\ion{Zn}{i} line only. Further, this is also the line that we rejected
from further analysis in Bensby et al.~(\cite{bensby}) since it was
suspected to have a hidden blend that was growing with
metallicity. For metallicities below [Fe/H]\,$\approx 0$ we, however,
found that the blend should have less influence. This is most likely 
the reason for the good agreement between our [Zn/Fe] trends.

Comparing our thin disk [Zn/Fe] trend with Reddy et al.~(\cite{reddy})
we see that in the range $\rm -0.4 \lesssim [Fe/H] \lesssim -0.2$ their stars
have [Zn/Fe] in the range $-0.1$\,dex to $+0.2$\,dex. This is higher than 
what we see for our thin disk stars that have [Zn/Fe] in the range 
$-0.1$\,dex to $0$\,dex in the same metallicity bin. Our thick disk stars 
have [Zn/Fe] in the range $0$\,dex to $+0.2$\,dex which means that by 
combining the [Zn/Fe] trends for our thin and thick disks 
we would see the same spread in [Zn/Fe] as Reddy et al.~(\cite{reddy}).
However, as we saw in Sect.~\ref{sec:errors}, there seem to be an offset of 
about $0.05--0.10$\,dex between our [Zn/Fe] abundances and those in 
Reddy et al.~(\cite{reddy}). Taking this into account will put our
thin disk [Zn/Fe] trend on the same level as the one in 
Reddy et al.~(\cite{reddy}), or vice versa. But, it will not give any 
insight into why the thin disk [Zn/Fe] trend  in Reddy et al.~(\cite{reddy}) 
show a larger scatter than what we see in our [Zn/Fe] trend. It is probably 
due to the analysis in which only
one or two spectral lines are analyzed, which unevitably leads to larger 
internal errors unless extreme care is taken.

%==============================================================================
\subsection{[$\alpha$/Fe] at different Z$_{\rm max}$}

%------------------------------------------------------------------------------
\begin{figure}
\resizebox{\hsize}{!}{
        \includegraphics[bb=18 144 592 445,clip]{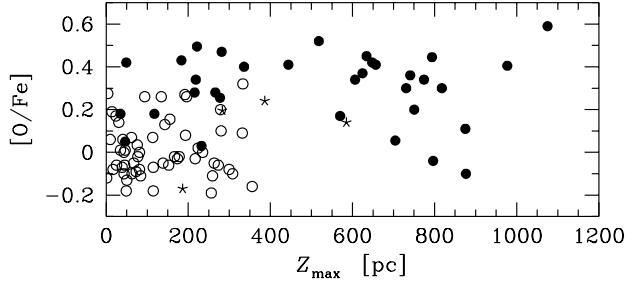}}
\caption{
        [O/Fe] versus $Z_{\rm max}$ for the 36 stars from this study
        and the 66 stars from Bensby et al.~(\cite{bensby}). Thin disk and 
        thick disk stars are 
        marked by empty and filled circles, respectively. Transition objects 
        are marked by asterisks.
         }
\label{fig:zmax}
\end{figure}
%------------------------------------------------------------------------------

%------------------------------------------------------------------------------
\begin{figure}
\resizebox{\hsize}{!}{
        \includegraphics[bb=18 144 592 718,clip]{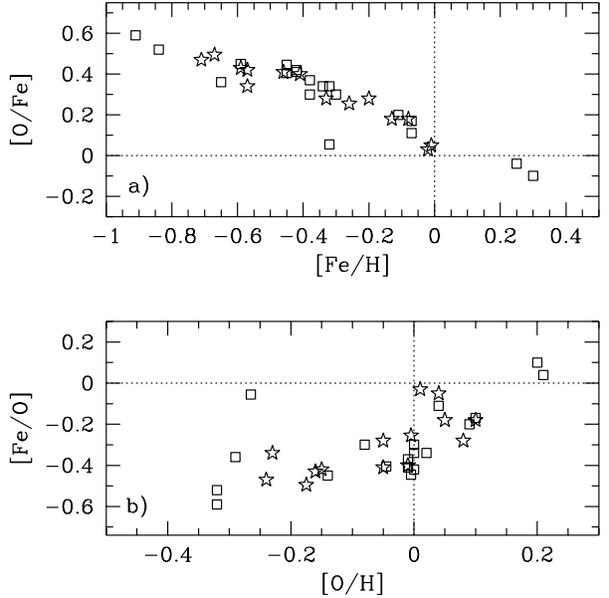}}
\caption{
        Abundance trends in the thick disk for oxygen. The thick disk
        sample has been divided into two
        sub-groups: stars that have $Z_{\rm max}>500$\,pc (open squares); 
        and stars that have $Z_{\rm max} \leq 500$\,pc (open stars).
         }
\label{fig:zmax2}
\end{figure}
%------------------------------------------------------------------------------

Since our thick disk stellar sample is far from complete and
is biased towards higher metallicities we can not use it to probe for
vertical gradients in the thick disk metallicity distribution. However, it
can be used to investigate if there are differences in the abundance trends
at various heights above the Galactic plane. If the trends are similar this
would indicate that the thick disk stars come from a stellar population 
that initially was homogeneous and well mixed. 

The maximum vertical distance ($Z_{\rm max}$) a star can reach above the plane
can be estimated from (L. Lindegren 2003, priv. comm.):
%------------------------------------------------------------------------------
\begin{equation}
        Z_{\rm max} = \sqrt{Z^{2} + (k\cdot W_{\rm LSR})^{2}},
\label{eq:zmax}
\end{equation}
%------------------------------------------------------------------------------
where
%------------------------------------------------------------------------------
\begin{equation} 
        k\,=\,(85\,{\rm Myr})\,/\,(2\,\cdot\,\pi)\,({\rm pc\,/\,Myr}) 
        \,/\,({\rm km\,/\,s}),
\end{equation}
%------------------------------------------------------------------------------
and $Z$ is the present distance of the star from the Galactic plane
(in pc) and $W_{\rm LSR}$ is its present velocity in the
$Z$-direction.  Equation~(\ref{eq:zmax}) assumes that stars in the
solar neighbourhood oscillates harmonically in the $Z$-direction with
a period of 85\,Myr, independent of their other motions. Since the
density of stars decreases with $|Z|$, Eq.~(\ref{eq:zmax}) will
underestimate $Z_{\rm max}$ for stars that have large $W_{\rm LSR}$
velocities. The relationship is however sufficient for our purposes.

%------------------------------------------------------------------------------
\begin{figure*}
\resizebox{\hsize}{!}{
        \includegraphics[bb=18 195 592 455,clip]{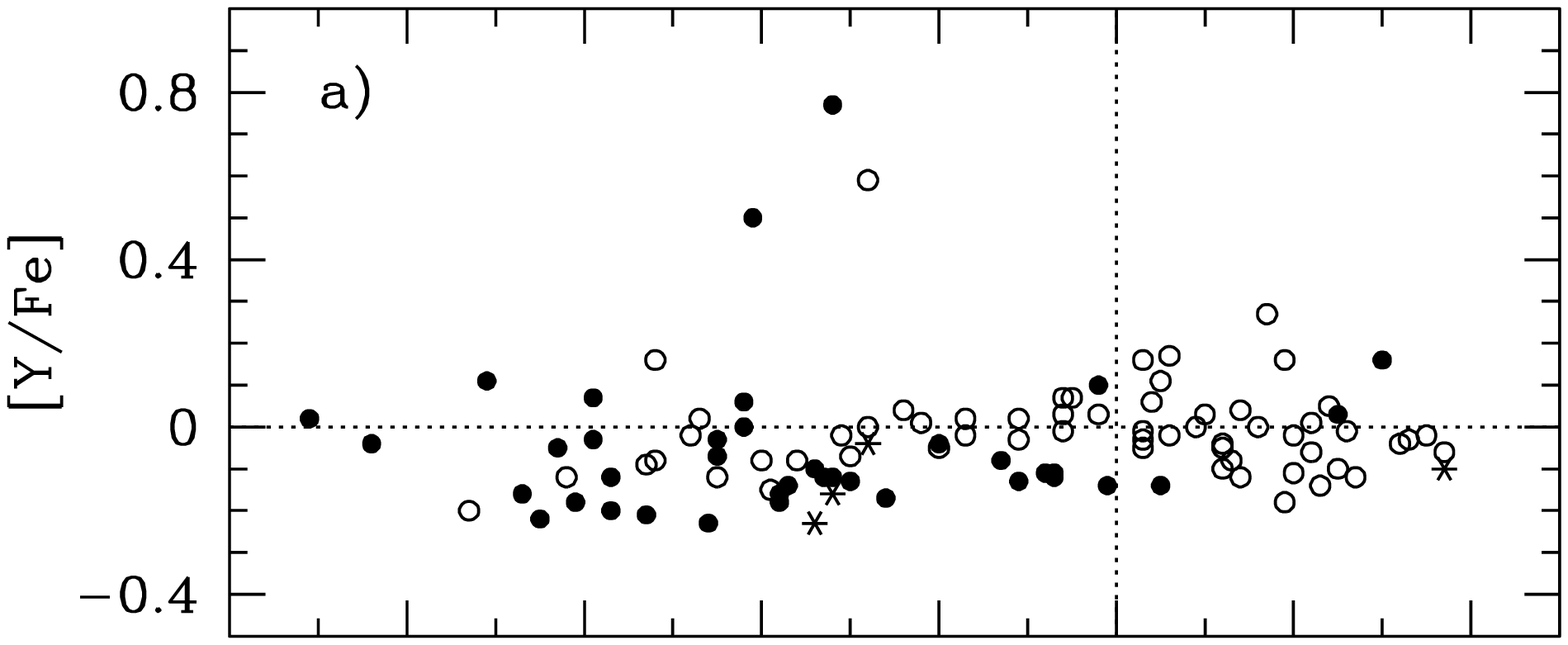} \hspace{10mm}
        \includegraphics[bb=18 195 592 455,clip]{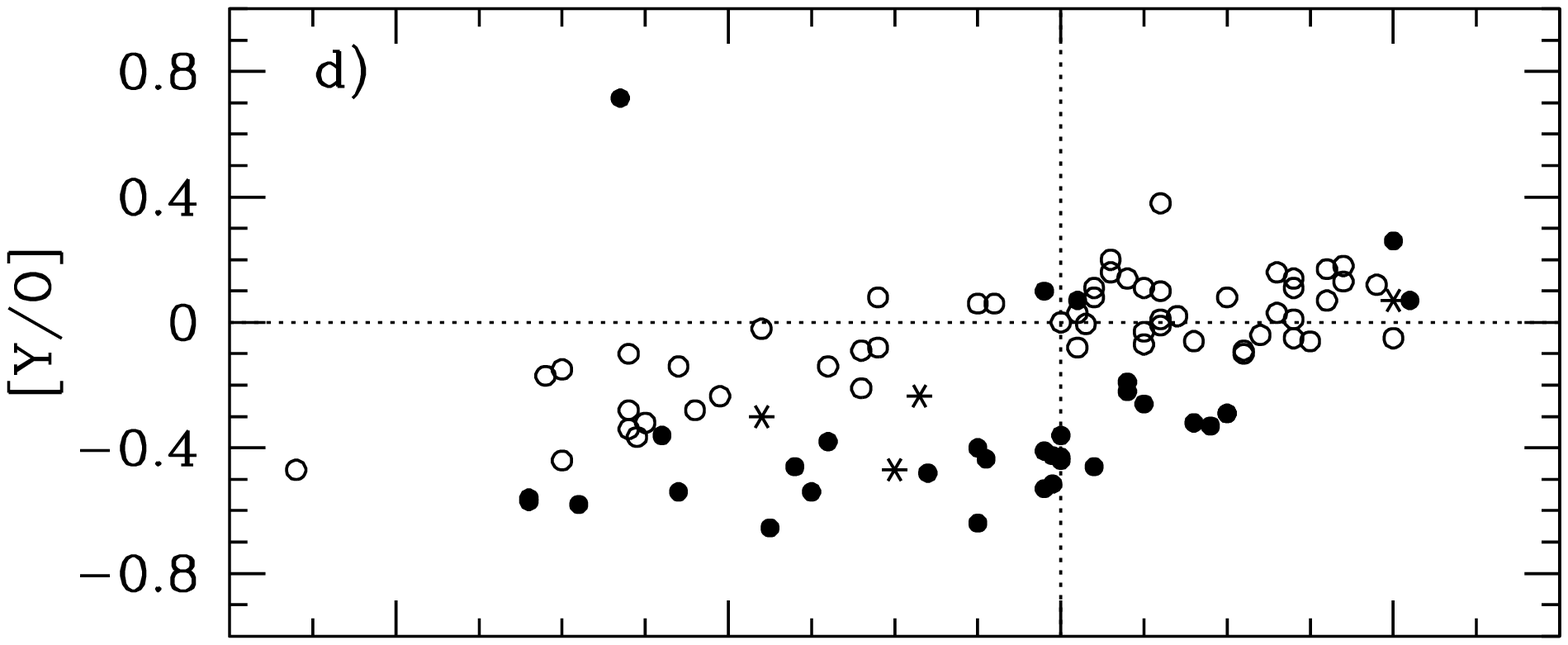}}
\resizebox{\hsize}{!}{
        \includegraphics[bb=18 195 592 450,clip]{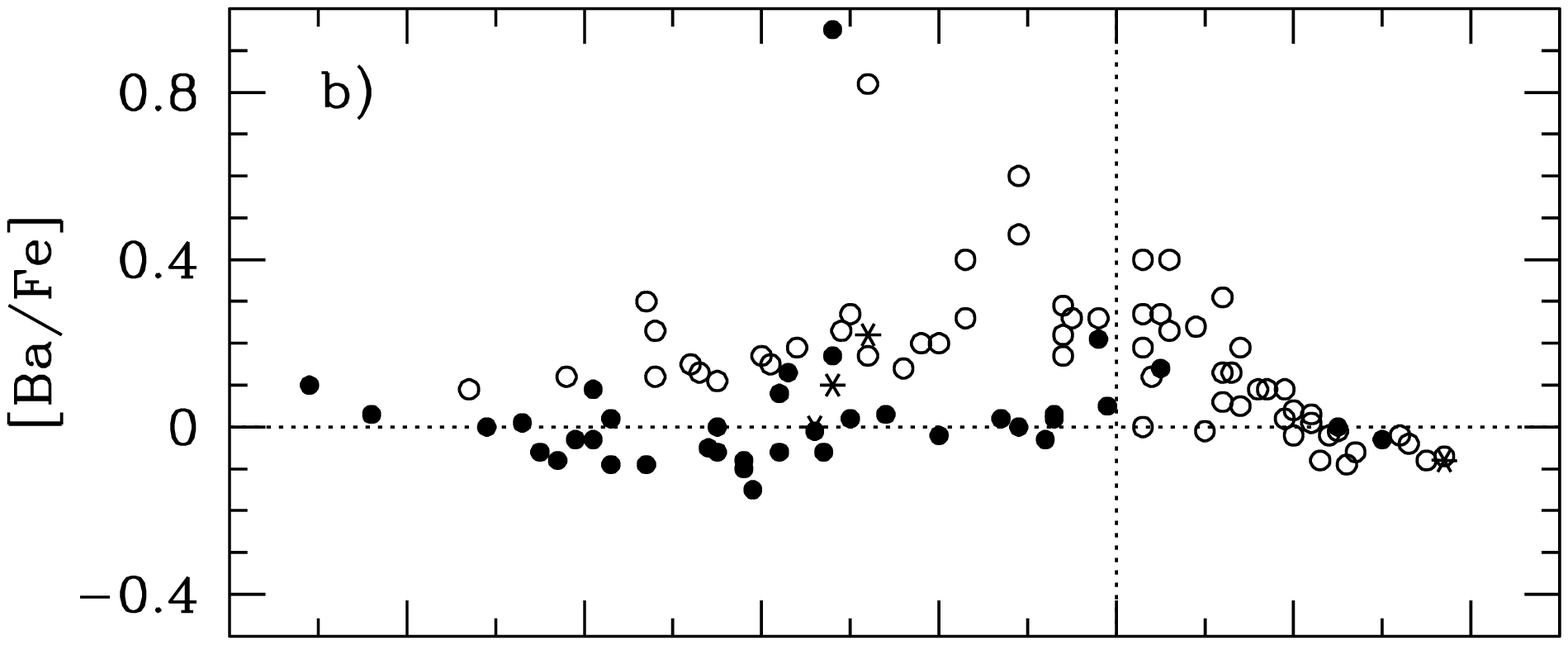} \hspace{10mm}
        \includegraphics[bb=18 195 592 450,clip]{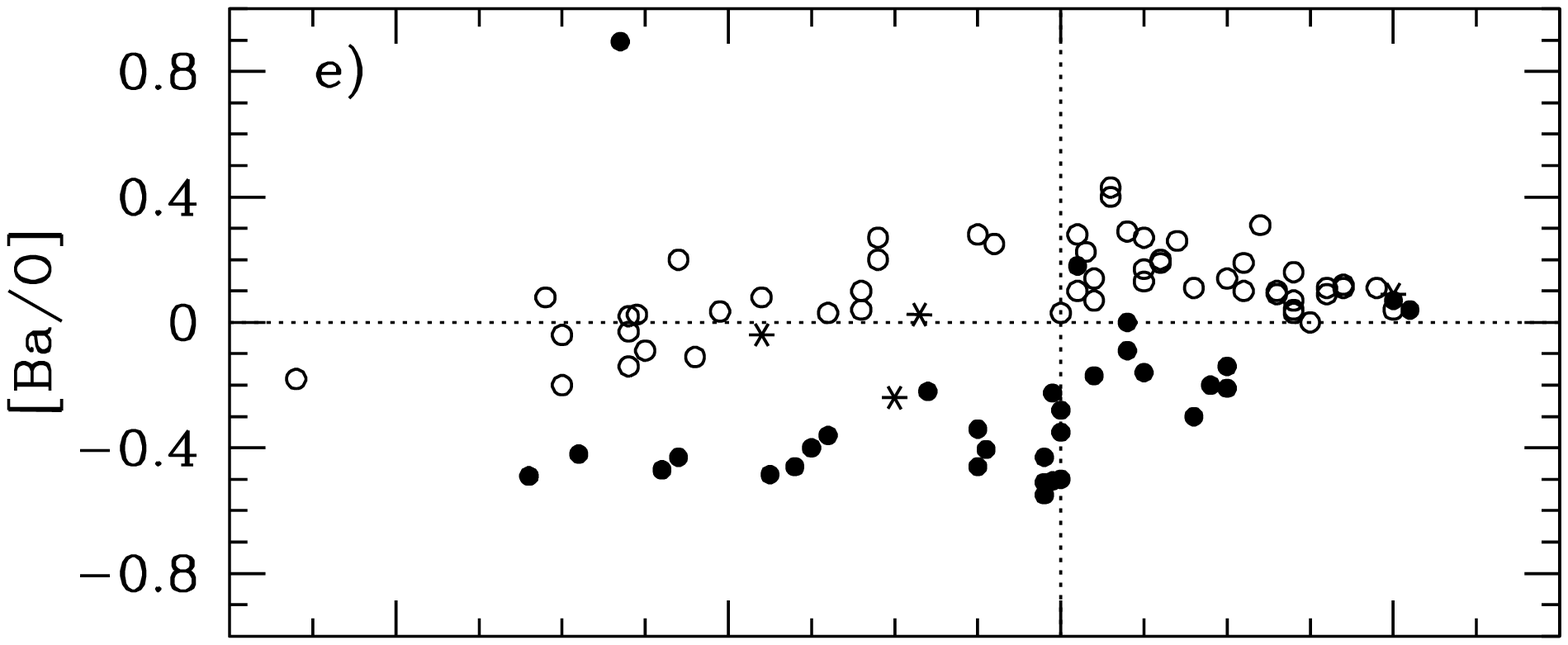}}
\resizebox{\hsize}{!}{
        \includegraphics[bb=18 144 592 450,clip]{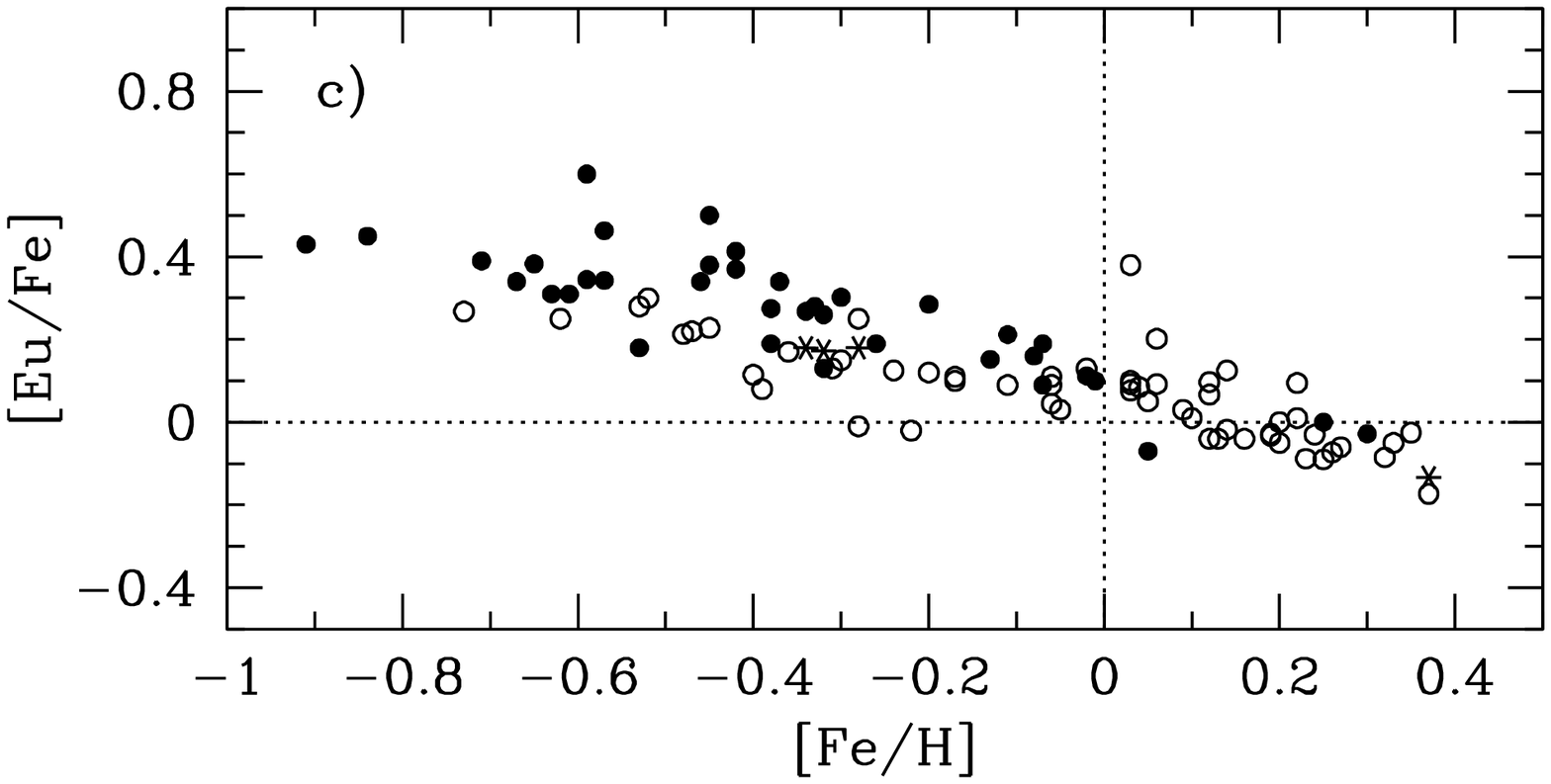} \hspace{10mm}
        \includegraphics[bb=18 144 592 450,clip]{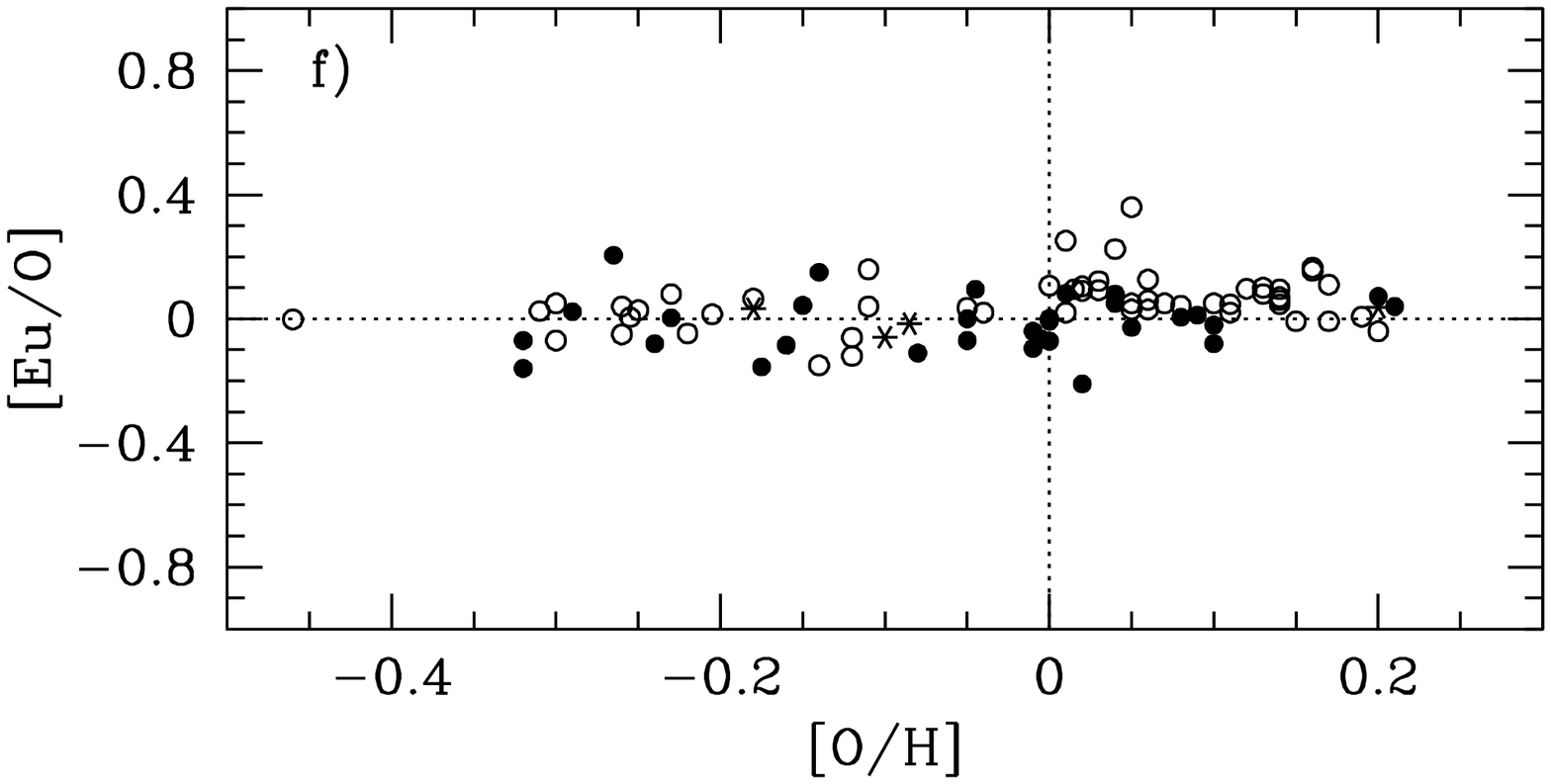}}
\caption{
        Elemental abundances relative to Fe. Dotted lines indicate
        solar values. Thin disk and thick disk stars are marked by
        empty and filled circles, respectively. Transition objects are
        marked by asterisks.
         }
\label{fig:rs_fe}
\end{figure*}
%------------------------------------------------------------------------------

Figure~\ref{fig:zmax} shows the [O/Fe] ratio as a function of 
$Z_{\rm max}$ for our sample.  The thin disk stars are all confined to 
within $\sim$\,300\,pc of the Galactic plane, whereas the thick disk stars 
move in orbits reaching  vertical distances up to 1\,kpc or
more. Dividing the thick disk sample at $\sim$\,500\,pc give two
sub-samples of approximately equal sizes.

As can be seen in Fig.~\ref{fig:zmax2} the [O/Fe] trends for the two
sub-samples are the same. An important point is that the "knee" is
located at the same [Fe/H] for both sub-samples.  If the Galactic
thick disk formed in a fast dissipational collapse, with a proposed
time scale of $\sim$\,400\,Myr (see, e.g., Burkert et
al. \cite{burkert}), it is likely that the position of the "knee"
would differ in the two sub-samples. SN\,II that have a time-scale of
typically 10\,Myr would then have time to enrich the interstellar
medium with even more of the $\alpha$-elements at higher [Fe/H] in the
thick disk sub-sample that formed closer to the Galactic plane.  The
invariance of our abundance trends with distance from the plane
instead indicates that the thick disk stellar population was well
mixed before it got kinematically heated. This can for 
example be accomplished if
the stars in a pre-existing old thin disk got kinematically heated by
the tidal interaction with a companion galaxy that either merged with,
or passed close by, the Galaxy.

%==============================================================================
\subsection{The $r$- and $s$-process element abundance trends}

Figure~\ref{fig:rs_fe}a--c shows our results for the $r$- and
$s$-process elements Y, Ba, and Eu with Fe as the reference element,
and Fig.~\ref{fig:rs_fe}d--f with oxygen as the reference element.

\paragraph{Eu:} Eu is supposed
to be an almost pure $r$-process element 
(approximately 94\,\%
$r$-process and 6\,\% $s$-process, Arlandini et al.~\cite{arlandini}) 
which
means that it mainly should come from SN\,II where the neutron flux is
sufficient for the $r$-process to occur (i.e. the neutron density is
so high that the neutron-capture timescale is much smaller than the
$\beta$-decay timescale).  The [Eu/Fe] trend in Fig.~\ref{fig:rs_fe}a
is very similar to the [O/Fe] trend in Fig.~\ref{fig:x_fe}a.  The
thick disk [Eu/Fe] trend also shows a turn-over at [Fe/H]\,$\approx
-0.4$, and the thin disk [Eu/Fe] trend shows the same shallow decline
over the whole range in [Fe/H]. The downward trend that we found for
[O/Fe] at [Fe/H]\,$>0$ is also present in [Eu/Fe] but with a larger
scatter.  The generally good agreement between Eu and oxygen, which is
further illustrated in Fig.~\ref{fig:rs_fe}f where we plot [Eu/O]
versus [O/H], indicates that these two elements indeed originate from
the same type of environments.

Our [Eu/Fe] trend for the thin disk is in good agreement with previous
studies (Woolf et al.~\cite{woolf}; Koch \& Edvardsson~\cite{koch};
Mashonkina \& Gehren~\cite{mashonkina}), and our thick disk [Eu/Fe]
trend is in agreement with Mashonkina \& Gehren~(\cite{mashonkina})
for the metallicities where their thick disk stars overlap with ours
(i.e. [Fe/H]\,$\lesssim -0.3$). The continuing 
decline that we see in [Eu/Fe] for
the thick disk for [Fe/H]\,$> -0.3$ is, on the other hand, new.

\paragraph{Ba and Y:} 
The s-process contributions to the solar composition
is for Ba 81\,\% (Arlandini et al.~\cite{arlandini}, but see also 
Travaglio et al.~\cite{travaglio}) and for Y 74\%
(Travaglio et al.~\cite{travaglio2}).  
In the
$s$-process the neutron flux is low, which means that the radioactive
isotopes will have time to $\beta$-decay between the neutron-captures.
Probable sites for the $s$-process are the atmospheres of stars on the
asymptotic giant branch (AGB stars) (e.g., Busso et al.~\cite{busso}).
If this is the case the enrichment of the $s$-process elements to the
interstellar medium probably occur on a timescale similar to elements
originating in SN\,Ia.

The [Y/Fe] and [Ba/Fe] trends are different for the thin and thick
disks (Fig.~\ref{fig:rs_fe}a and b).  Especially [Ba/Fe] is distinct
and well separated for the two disks.  For the thick disk stars the
[Ba/Fe] trend (Fig.~\ref{fig:rs_fe}b) is flat, lying on a solar
ratio. [Y/Fe] for the thick disk shows a larger scatter and has a flat
appearance with underabundances between 0 and $-0.2$\,dex
(Fig.~\ref{fig:rs_fe}a).  The thin disk [Ba/Fe] trend shows a
prominent rise from the lowest [Fe/H] until reaching solar
metallicities, after which it starts to decline. The [Y/Fe] trend for
the thin disk is similar but shows a considerably larger scatter and
not such a well-defined trend as that for [Ba/Fe].

%==============================================================================
\subsection{Evolution of Ba and Eu}

%------------------------------------------------------------------------------
\begin{figure}
\resizebox{\hsize}{!}{
        \includegraphics[bb=18 144 592 718,clip]{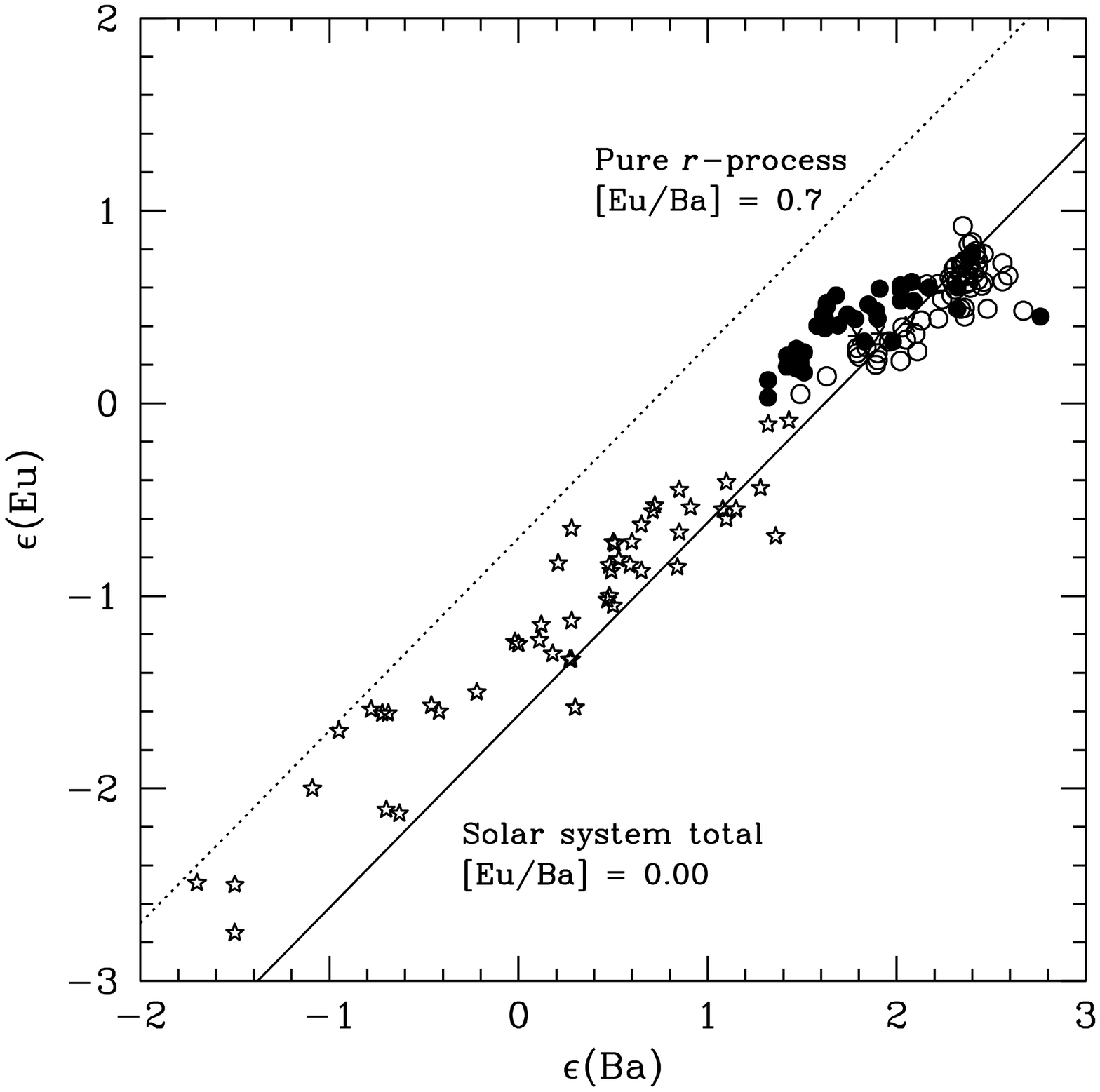}}
\caption{
        An illustration of the relative contributions from the $r$-
        and $s$-process to the elements Eu and Ba. The dotted line
        shows the pure $r$-process contribution to Eu and Ba while the
        full line shows the solar system mix of $r$- and $s$-process
        contributions (Arlandini et al.~\cite{arlandini}).  Thin and
        thick disk stars are marked by open and filled circles,
        respectively. Transition objects are marked by asterisks
        ($\ast$), and 'open stars' data from Burris et al.~(\cite{burris}).
         }
\label{fig:eps_ba_eu}
\end{figure}
%------------------------------------------------------------------------------
%------------------------------------------------------------------------------
\begin{figure}
\resizebox{\hsize}{!}{
           \includegraphics[bb=18 145 592 490,clip]{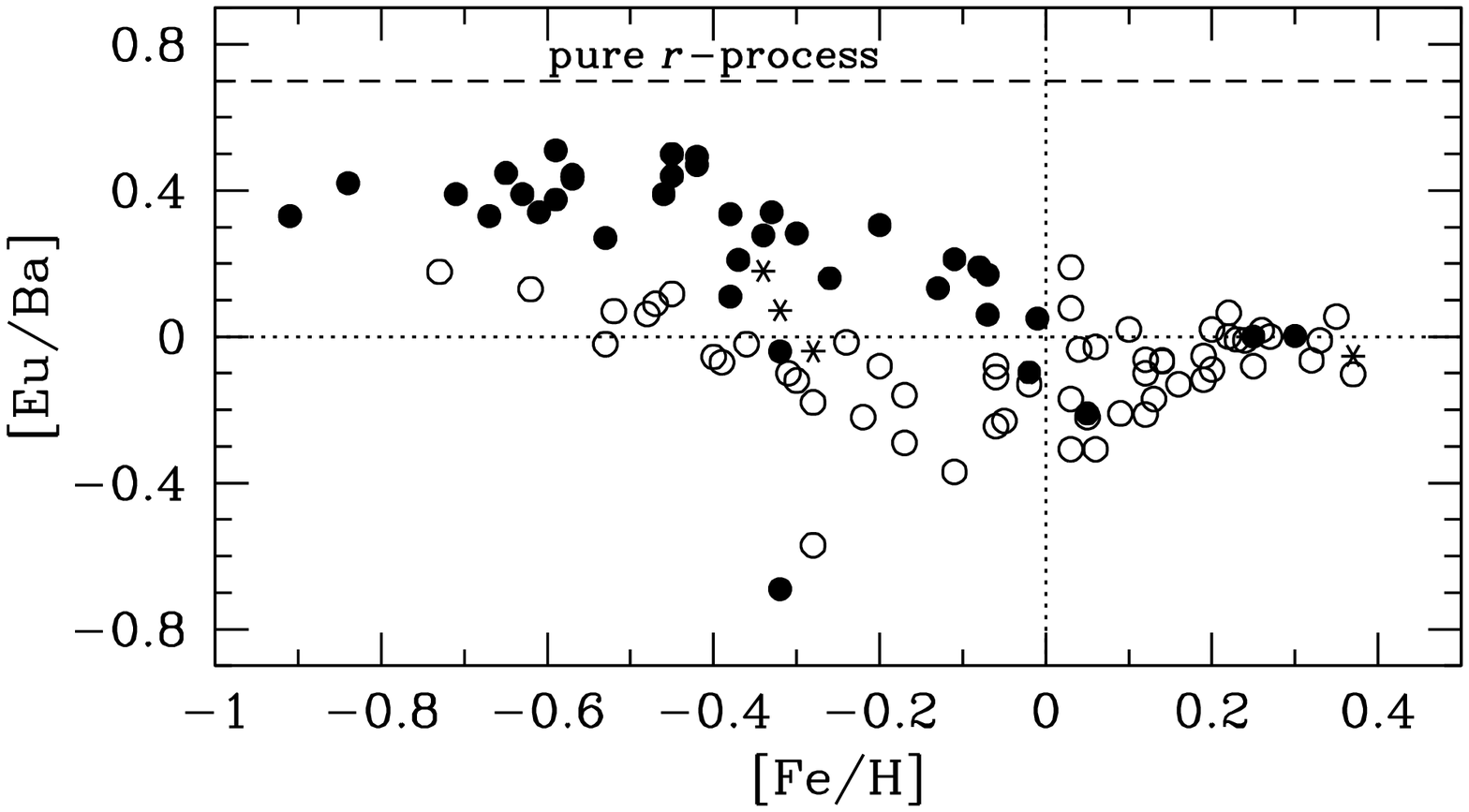}}
\caption{
        [Eu/Ba] versus [Fe/H]. The dashed lines show the 
        pure r-process ratio which is [Eu/Ba]\,=\,0.7
        and has been calculated from the yields given in 
        Arlandini et al.~(\cite{arlandini}).
         }
\label{fig:euba_euy}
\end{figure}
%------------------------------------------------------------------------------

In Fig.~\ref{fig:eps_ba_eu} we compare our Eu and Ba data with
predictions for pure $r$-process composition ($\rm [Eu/Ba]=0.7$, calculated
from the yields given in Arlandini et al.~\cite{arlandini}) 
and a solar mixture  of $r$- and $s$-process contributions 
(i.e., $\rm [Eu/Ba]=0$).  Also included in this plot are low-metallicity 
halo stars (giants) from Burris et al.~(\cite{burris}). The
thin disk shows a solar system mix for all metallicities while the
thick disk has not yet experienced the full contribution of
$s$-processed material from low mass AGB stars, i.e., it is closer to
the pure $r$-process line.

A first tentative interpretation of these results is that star
formation went on long enough in the thick disk so that AGB stars
started to contribute to the chemical enrichment, but only just long
enough that a solar system mix was reached. After the formation of
stars stopped in the thick disk the remaining gas settled into a new
thinner disk. Most likely, fresh material of lower metallicity was
accreted before star formation started in what is today's thin disk.
The relative $r$- and $s$-process contributions will not change by
this dilution if the infalling material is pristine, so in this case
the first thin disk stars to form will retain the mixture that was at
the end of star formation in the thick disk. The absolute abundances
of Ba and Eu in the thin disk will, however, be shifted towards lower
values. If on the other hand, the infalling material has experienced
enrichment of $r$- and/or $s$-process elements the mixture should
change.  Our data appear to indicate that this has not been the case
and hence that the infalling material was most likely primordial.

This is further illustrated in Fig.~\ref{fig:euba_euy} 
where we plot [Eu/Ba] versus [Fe/H] for our stellar sample only. 
The most metal-rich thick disk stars (at $\rm [Fe/H]\approx 0$) and 
the most metal-poor thin disk stars (at $\rm [Fe/H]\approx -0.7$) have
approximately the same [Eu/Ba] ratio. So while pristine material falls
into the disk the gas (from which the thin disk stars form) gets more 
metal-poor, the [Eu/Ba} ratio is preserved.

%==============================================================================
\subsection{Nucleosynthesis in low mass AGB stars}

\begin{figure}
\resizebox{\hsize}{!}{
        \includegraphics[bb=18 144 592 600,clip]{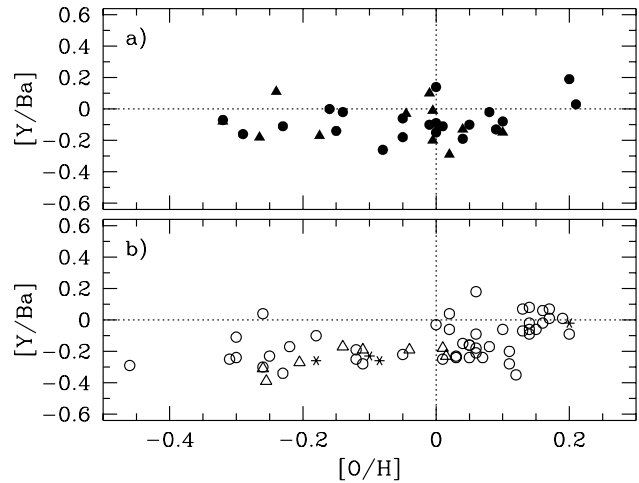}}
\caption{
        [Y/Ba] versus [O/H] for the {\bf a)} thick disk stars
        and in {\bf b)} for the thin disk stars (with the four ``transition"
        objects included as well).
         }
\label{fig:y_ba_o}
\end{figure}

An interesting result is found when studying the trend of [Y/Ba]
versus [O/H], see Fig.~\ref{fig:y_ba_o}. For both the thin and thick
disks this trend is first flat but after solar metallicity a gentle
upward trend is seen. This could be explained as a metallicity effect
in AGB nucleosynthesis (Busso et al.~\cite{busso2}).  Figure~1 in
Busso et al.~(\cite{busso2}) shows how the relative production of
light (e.g. Y) and heavy (e.g. Ba) $s$-process elements change as a
function of metallicity. For metallicities below solar the [Y/Ba]
ratio is roughly flat, i.e. Y and Ba are produced in the same ratio in
the low mass AGB stars. However, around, or slightly above, solar
metallicity this balance changes such that the lighter $s$-process
elements are favoured over the heavy $s$-process elements. Hence
[Y/Ba] increases.
%------------------------------------------------------------------------------
%------------------------------------------------------------------------------

%==============================================================================
\subsection{Transition objects} \label{sec:transition}

In our stellar sample we have four stars whose kinematical properties
lie in between the definitions of the thin and thick disk populations
that we have used (see Table~\ref{tab:kinematics}).  Depending on the
choice of the normalization of the thick disk density in the solar
neighbourhood they will alter their classifications as either thin
disk or thick disk stars (see discussion in the Appendix).  When
looking at their chemical compositions we see that they not only are
intermediate in terms of kinematics but also in terms of abundances at
a given [Fe/H] (see Figs.~\ref{fig:x_fe} and \ref{fig:rs_fe}).  Which
stellar population these stars should belong to, or if they form a
distinct population by their own, can only be investigated with a
larger sample of such stars.

%==============================================================================
\subsection{Deviating stars - outliers}  \label{sec:outliers}

From our abundance plots it is evident that not all stars follow the
trends as outlined by the majority of stars in the two disk
populations.  There are, in particular, three stars (HIP~2235,
HIP~15510, and HIP~16788) that seem to show suspiciously high
abundances in either or both of Y and Ba (see Figs.~\ref{fig:rs_fe}a,
b, d, and e). To enable a direct comparison we list their chemical
properties in Table~\ref{tab:outliers} and discuss each of them in
turn. \\
{\bf HIP 2235} is a thin disk star and has spectral type F6V (according
to the Simbad database) and shows high over-abundances in both Ba and Y 
that is not recognized in any of the other thin disk stars (or the thick 
disk stars). This s-enhancement can be due to that s-enriched material has 
been transferred from from a companion star into the stellar atmosphere. 
The companion star could have been an AGB star, in which these elements are 
believed to be synthesized (see, e.g., Abia et al.~\cite{abia}), that now 
is invisible since  it has evolved into a white dwarf. \\ 
{\bf HIP 15510} is a G8V thick disk star and show an abnormal enhancement
in Y but not in Ba. It is not unlikely that this star not has been
subject to the same type of mass-transfer as HIP~2235 maybe has
been. It is namely well possible to have s-stars with Y-enhancement
and no Ba-enhancement and vice versa (see, e.g., review by Busso et
al.~\cite{busso3}). However, the Y abundance for HIP~15510 is based on
one spectral line only, making it highly uncertain. \\
{\bf HIP 16788} is a G0 thick disk star that has no luminosity classification
in the Simbad database. However, from our derived $\log g = 4.24$ it
is most likely also a main sequence star. It is highly enhanced in both
Y and Ba as in the case for HIP~2235.

Stars from the ``old sample'' that show deviating $\alpha$-abundances
were discussed in Sect.~9.4 in Bensby et al.~(\cite{bensby}) to which
the reader is referred since none of the new stars showed deviating
abundances for these elements.
%------------------------------------------------------------------------------
\begin{table}[t]
\centering
\caption{
         Abundances for deviating stars. Each star has three columns; 
         abundance; line-to-line scatter (1$\sigma$ standard deviation); 
         and (in parenthesis) number of lines that were used to derive 
         the abundance.
        }
\centering %\scriptsize
\setlength{\tabcolsep}{1.3mm}
\begin{tabular}{crlrlrl}
\hline \hline\noalign{\smallskip}
             &  \multicolumn{2}{c}{HIP 2235}  
             &  \multicolumn{2}{c}{HIP 15510}  
             &  \multicolumn{2}{c}{HIP 16788}  \\
             &  \multicolumn{2}{c}{Thin disk}
             &  \multicolumn{2}{c}{Thick disk}
             &  \multicolumn{2}{c}{Thick disk} \\
\noalign{\smallskip}
\hline\noalign{\smallskip}
\noalign{\smallskip}
$\rm [Fe/H]$ & $-0.28$ & 0.07 (65)  &  $-0.41$ & 0.08 (69) &  $-0.32$ & 0.07 (78) \\
$\rm [O/H]$  &         &            &  $-0.01$ & 0.00 (1)  &  $-0.27$ & 0.00 (1)  \\
$\rm [Eu/H]$ & $-0.19$ & 0.00 (1)   &          &           &  $-0.06$ & 0.00 (1)  \\
$\rm [Ba/H]$ &    0.54 & 0.06 (4)   &  $-0.56$ & 0.08 (3)  &    0.63  & 0.06 (4)  \\
$\rm [Y/H]$  &    0.31 & 0.14 (3)   &    0.09  & 0.00 (1)  &    0.45  & 0.12 (4)  \\
\noalign{\smallskip}
\hline
\end{tabular}
\label{tab:outliers}
\end{table}
%------------------------------------------------------------------------------

%==============================================================================
\section{Discussion and summary} \label{sec:summary}

In this study we have presented a differential abundance analysis
between the Galactic thin and thick disks for 14 elements (O, Na, Mg,
Al, Si, Ca, Ti, Cr, Fe, Ni, Zn, Y, Ba, and Eu) for a total of 102 nearby
F and G dwarf stars (including the stars from our previous studies
Bensby et al.~\cite{bensby}, \cite{bensby_syre}).
The results from the 36 stars in the new sample further
confirms, strengthens, and extends the results presented
in Bensby et al.~(\cite{bensby}, \cite{bensby_syre}).  
Results that are new in this study are those for the $r$- and
$s$-process elements Y, Ba, and Eu, where we find the thin and thick
disks abundance trends to be distinct and well defined. We also see
indications of a metallicity effect in the AGB nucleosynthesis of Y
and Ba, such that Y is favoured over Ba at higher [Fe/H]. Our results
for Eu show that Eu abundances follow the oxygen abundances very
well. This confirms that Eu is an element that mainly is produced in
SN\,II.

In our studies we have included thick disk stars with [Fe/H]\,$\gtrsim-0.35$,
which no other study have.
At these higher metallicities we find that the [$\alpha$/Fe]
trends, at [Fe/H]\,$\approx-0.4$, turns over and decline towards solar
values where they merge with the thin disk [$\alpha$/Fe] trends.  The
observed down-turn (or ``knee") in the thick disk [$\alpha$/Fe] trends
at [Fe/H]\,$\approx -0.4$ can be interpreted as a signature of the
contribution from SN\,Ia to the chemical enrichment of the stellar
population under study. Massive stars
($M\gtrsim10$\,$\mathcal{M}_{\sun}$) explode as core-collapse
supernovae type II (SN\,II) and enrich the interstellar medium with
$\alpha$-elements and lesser amounts of heavier elements such as the
iron peak elements (e.g. Tsujimoto et al.~\cite{tsujimoto}; Woosley \&
Weaver~\cite{woosley}).  Due to the short lifetimes of these massive
stars they enrich the interstellar medium in the early phases of the
chemical evolution and produce high [$\alpha$/Fe] ratios at the lower
metallicities.  SN\,Ia disperse large amounts of iron-peak elements
into the interstellar medium and none or little of $\alpha$-elements.
Since their low-mass progenitors are expected to have much longer
lifetimes than the SN\,II progenitors (e.g. Livio~\cite{livio}) there
will be a delay in the production of Fe as compared to the
$\alpha$-elements. Hence, when SN\,Ia start to contribute to the
enrichment, the [$\alpha$/Fe] ratios will decrease.

The fact that we see the signatures from SN\,Ia in the thick disk thus
means that star formation must have continued in the thick disk for a
time that was at least as long as the time-scale for SN\,Ia.  The
time-scale for a single SN\,Ia is very uncertain (see
e.g. Livio~\cite{livio}). However, we have seen from a study of ages
and metallicities in the thick disk that it has taken about
2--3\,billion years for the thick disk stellar population to reach a
metallicity of [Fe/H]\,$= -0.4$ (Bensby et al.~\cite{bensby_amr}).
Thus we can tentatively conclude that the SN\,Ia rate peaked at
$\sim2$--3 billion years from the start of the star formation in the
population that we today associate with the thick disk. The
age-metallicity relation in the thick disk that we find in that study
also indicates that star formation might have continued for 2--3
billion years after the peak in the SN\,Ia rate in order to reach
roughly solar metallicities.  The most important conclusion from this
is that the thick disk most probably formed during an epoch spanning
several ($>2$--3) billion years.  Data from solar neighbourhood stars
have also shown the SN\,Ia time-scale to be as long as
$\sim$\,1.5\,Gyr (Yoshii et al.~\cite{yoshii}).

We are able to draw further conclusions about the origin and chemical
evolution of the thick disk.  The observational constraints for a
formation scenario of the thick disk are:
\begin{enumerate}
\item   Distinct, smooth, and separated abundance trends between the thin and 
        thick disks.
\item   At a given [Fe/H] below solar metallicities the thick disk stars 
        are more enhanced in their $\alpha$-element abundances than the thin 
        disk
\item   The thick disk $\alpha$-element trends show signatures of
        enrichment from SN\,Ia.
\item   The abundance trends are identical for thick disk stars
        that reach different heights above the Galactic plane 
        ($>500$\,pc and $<500$\,pc).
\item   The thick disk stars have an older mean age than the thin disk stars
\item   AGB stars have contributed to the chemical enrichment of the thick disk
         but not as much as to the enrichment of the thin disk
\end{enumerate}
\begin{itemize}
\item[7.]
A less well established constraint from this study is that 
there might be a possible age-metallicity relation in the thick disk
indicating that star formation might have continued for several 
billion years.
\end{itemize}

Taking these constraints into consideration we argue that the currently most
probable formation scenario for the thick disk is an ancient merger
event between the Milky Way and a companion galaxy. In this event the
stellar population of the thin disk that was present at that time got
kinematically heated to the velocity distributions and dispersions
that we see in today's thick disk. We note that recent models of hierarchical
galaxy formation might be able to succesfully reproduce thick disks
in Milky Way like galaxies and the abundance trends might be 
fully explainable also in these models  (Abadi et al.~\cite{abadi}).

How can we explain the trends observed in the thin disk?
The thin disk stars, on average, are younger than the thick
disk stars. However, the low-metallicity tail in the metallicity distribution
of the thin disk stars overlap with the metallicity distribution
of the thick disk stars. A possible
scenario would be that once star
formation in the thick disk stops, there is a pause in the
star formation. During this time in-falling fresh gas
accumulates in the Galactic plane, forming a new thin disk.
Also, if there is any remaining gas from the thick disk it will settle
down onto the
new disk. Once enough material is collected, star formation is restarted in
the new thin disk. The gas, though, has been diluted by the metal-poor
in-falling gas. This means that the first stars to form in the thin disk will
have lower metallicities than the last stars that formed in the thick disk.

%==============================================================================
\begin{acknowledgements}

We would like to thank the developers of the Uppsala MARCS code, Bengt
Gustafsson, Kjell Eriksson, Martin Asplund, and Bengt Edvardsson who
we also thank for letting us use the {\it Eqwidth} abundance program.
Bj\"orn Stenholm is thanked for helping out with part of the
observations on La Palma, and we also thank our referee Roberto Gallino
for valuable comments that improved the analysis and text of the paper.  
This research has made
use of the SIMBAD database, operated at CDS, Strasbourg, France.

\end{acknowledgements}

%==============================================================================
\appendix

%==============================================================================
\section{Kinematical criteria for selecting thick disk stars in the solar 
neighbourhood} 
        \label{sec:appendix}

When selecting thin and thick disk stars we assume that the Galactic
space velocities ($U_{\rm LSR}$, $V_{\rm LSR}$, and $W_{\rm LSR}$) for the
thin disk, thick disk, and stellar halo have
Gaussian distributions. 
The space velocities
$U_{\rm LSR}$, $V_{\rm LSR}$, and $W_{\rm LSR}$ were calculated using our
measured radial velocities and positions, proper motions, and parallaxes from
the Hipparcos catalogue
(see Eqs.~A.1\,--\,A.4 in Bensby et al.~\cite{bensby}).
For each star we then 
calculate the probabilities that it belong to either the thin disk ($D$),
thick disk ($TD$), or the halo ($H$). By also take the fraction of 
thick disk stars in the solar neighbourhood into account, the final 
relationship for calculating the individual probabilities  are 
(see also Bensby et al.~\cite{bensby}):
%------------------------------------------------------------------------------
\begin{equation}
        P = X \cdot k \cdot \exp\left(
        -\frac{U^{2}_{\rm LSR}}{2\,\sigma_{\rm U}^{2}}
        -\frac{(V_{\rm LSR} - V_{\rm asym})^{2}}{2\,\sigma_{\rm V}^{2}}
        -\frac{W^{2}_{\rm LSR}}{2\,\sigma_{\rm W}^{2}}\right),
\label{eq:probabilities}
\end{equation}
%------------------------------------------------------------------------------
where $P$ is either $D$, $TD$ or $H$; and
%------------------------------------------------------------------------------
\begin{equation}
        k = \frac{1}{(2\pi)^{3/2}\,\sigma_{\rm U}
                                 \,\sigma_{\rm V}
                                 \,\sigma_{\rm W}},
\end{equation}
%------------------------------------------------------------------------------
normalizes the expression;  $\sigma_{\rm U}$, $\sigma_{\rm V}$,
$\sigma_{\rm W}$ are the characteristic velocity dispersions; $V_{\rm asym}$
is the asymmetric drift; and $X$ is the observed fraction of stars in the
solar neighbourhood for each population.
We calculated for each star the
``relative probabilities" $TD/D$ and $TD/H$. Thin disk stars were selected as
those with $TD/D<0.1$ (i.e. at least ten times more probable of being a
thin disk stars than a thick disk star) and thick disk stars as those with
$TD/D>10$ (i.e. at least ten times more probable of being a
thick disk stars than a thin disk star). $TD/H>1$ was required for both
the thin and thick disk stars.
%------------------------------------------------------------------------------
\begin{figure}
\resizebox{\hsize}{!}{\includegraphics{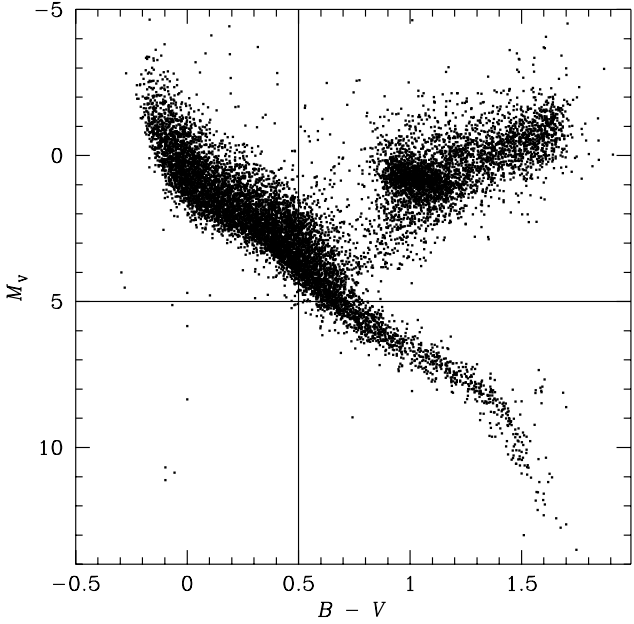}}
\caption{
        CMD for the full stellar sample (12\,634 stars).
         }
\label{fig:cm_all}
\end{figure}
%------------------------------------------------------------------------------

Of the parameters that has the largest influence on the
derived probability ratios, the normalization factor
is the one that is the least well constrained.
In Bensby et al.~(\cite{bensby}) we used a value of 6\,\%. 
The lowest value 2\,\% was found by Gilmore \& Reid~(\cite{gilmore}) and
Chen~(\cite{chen2}). Intermediate values $\sim$\,6\,\% were found by
Robin et al.~(\cite{robin}) and Buser et al.~(\cite{buser}), and higher values
$\sim$\,15\,\% by Chen et al.~(\cite{chen3}) and
Soubiran et al.~(\cite{soubiran}).

We will here show that an increase of the local normalization
of the thick disk stars from 6\,\% to 10\,\% (and consequently a lowering of
the thin disk density from 94\,\% to 90\,\%) is motivated.
Changing to 10\,\% will not influence the thick disk sample in
Bensby et al.~(\cite{bensby}) since the only effect is to raise the $TD/D$
ratios by a factor of $\sim$\,1.7 (see column 2 in Table~\ref{tab:tdd}).
Instead those thick disk stars will have their classicications
strengthened. The thin disk sample in Bensby et al.~(\cite{bensby})
will not change either since all those stars had their $TD/D$ ratios
well below 0.1 (typically 0.01).

We select our stars from the same data set as in 
Feltzing et al.~(\cite{feltzing}) and Feltzing \& Holmberg~(\cite{feltzing2}). 
In brief this includes all stars in the
Hipparcos catalogue (ESA~\cite{esa}) that have relative errors in their
parallaxes less than 25\,\% and that have published radial velocities
(see e.g. Bensby et al.~\cite{bensby}; Feltzing \& Holmberg~\cite{feltzing2}).
This sample consists of 12\,634 stars. Note that known binaries have
been excluded from the data (see Feltzing et al.~\cite{feltzing}).

%-----------------------------------------------------------------------------
\begin{table}
\centering
\caption{
        Characteristic velocity dispersions
        ($\sigma_{\rm U}$, $\sigma_{\rm V}$, and $\sigma_{\rm W}$) in the thin
        disk, thick disk, and stellar halo, used in
        Eq.~(\ref{eq:probabilities}).
        $V_{\rm asym}$ is the asymmetric drift.
        }
\begin{tabular}{lcccr}
\hline \hline\noalign{\smallskip}
        & $\sigma_{\rm U}$
        & $\sigma_{\rm V}$
        & $\sigma_{\rm W}$
        & $V_{\rm asym}$ \\
\noalign{\smallskip}
        & \multicolumn{4}{c}{----------~~[km\,s$^{-1}$]~~----------}     \\
\noalign{\smallskip}
\hline\noalign{\smallskip}
   Thin disk ($D$)   & $~~35$  & 20    & 16    & $-15$    \\
   Thick disk ($TD$) & $~~67$  & 38    & 35    & $-46$    \\
   Halo ($H$)        & $160$   & 90    & 90    & $-220$   \\
\hline
\end{tabular}
\label{tab:dispersions}
\end{table}
%-----------------------------------------------------------------------------
%------------------------------------------------------------------------------
\begin{table}
\centering
\caption{
        The number of stars in given $TD/D$ intervals for
        different values on the local density of thick disk
        stars ($X_{\rm TD}$). The second column indicates the factor by
        which the $TD/D$ ratios change when varying the normalization
        (with the 10\,\% density as base). The corresponding CM-diagrams
        can be seen in Fig.~\ref{fig:cm_many}.
        }
\centering %\scriptsize%\tiny
\begin{tabular}{rlrrrrrr}
\hline \hline\noalign{\smallskip}
      $X_{\rm TD}$
   &
   &  \multicolumn{4}{c}{---------------~~$N_{\rm stars}$~~---------------} \\
\noalign{\smallskip}

        &
        &  $<$\,0.1
        &  0.1\,--\,1
        &  1\,--\,10
        &  $>$\,10  \\
\hline
\noalign{\smallskip}
2\%     & 0.2  & 11781 &  362 & 166 & 261 \\
6\%     & 0.6  & 11305 &  689 & 238 & 347 \\
10\%    & 1    & 10969 &  946 & 285 & 383 \\
14\%    & 1.5  & 10623 & 1215 & 327 & 421 \\
\noalign{\smallskip}
\hline
\end{tabular}
\label{tab:tdd}
\end{table}
%------------------------------------------------------------------------------
%------------------------------------------------------------------------------
\begin{figure*}
\resizebox{\hsize}{!}{\includegraphics[bb=40 17 632 779,clip]{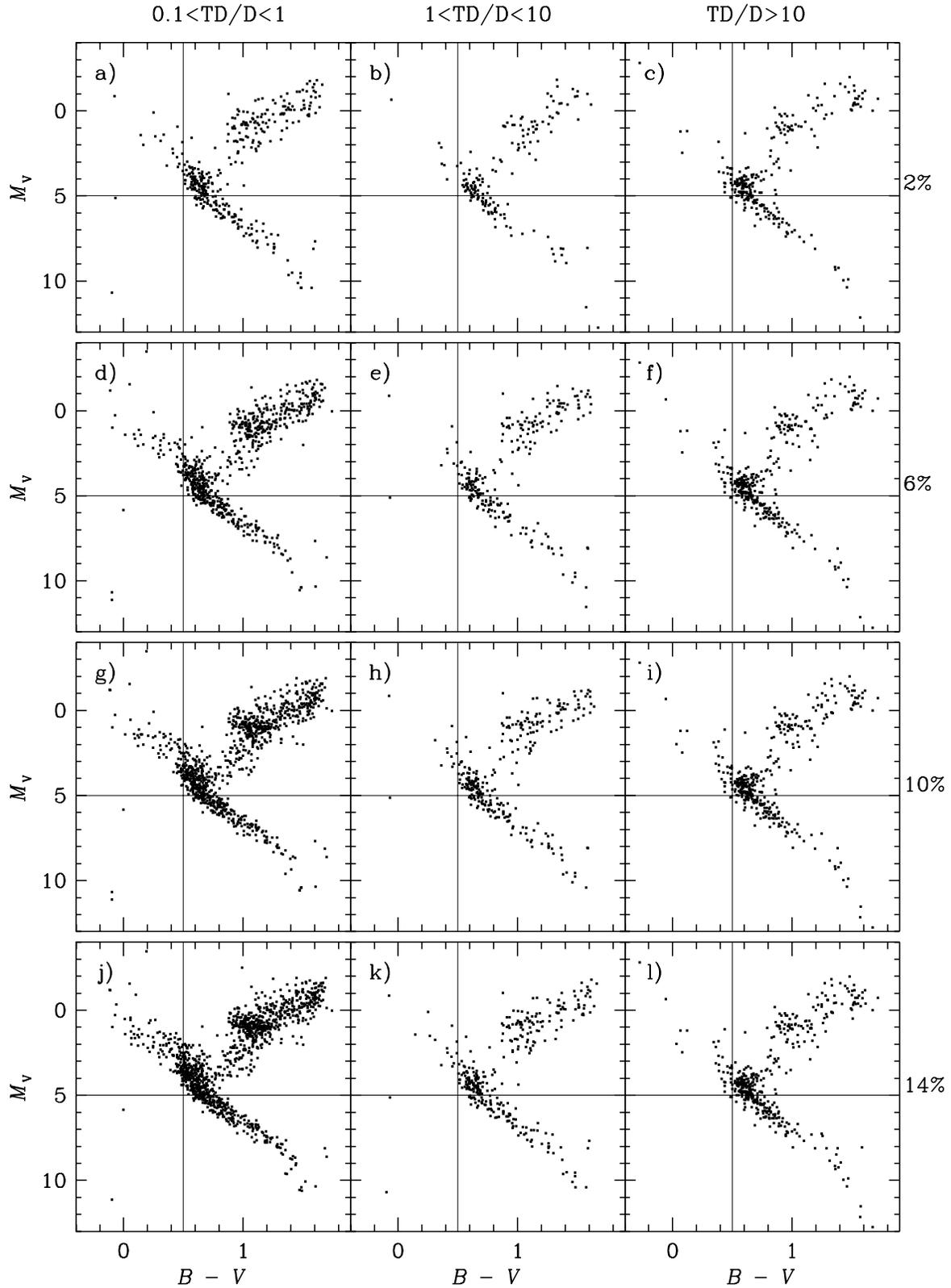}}
\caption{
        CMDs for different interval of $TD/D$ when changing the
        local density of thick disk stars (2\,\%, 6\,\%, 10\,\%, and 14\,\%).
        The number of stars in each CMD is given in Table~\ref{tab:tdd}.
         }
\label{fig:cm_many}
\end{figure*}
%------------------------------------------------------------------------------

The colour-magnitude diagram (CMD) for all 12\,634 stars is shown in
Fig.~\ref{fig:cm_all}. Figure~\ref{fig:cm_many} show the CMD for three
different $TD/D$ intervals: $0.1<TD/D<1$ (i.e ``low probability" thin
disk stars); $1<TD/D<10$ (i.e ``low probability" thick disk stars);
and $TD/D>10$ (i.e ``high probability" thick disk stars), with four
different values of the thick disk normalization in the solar neighbourhood
(2\,\%, 6\,\%, 10\,\%, 14\,\%). The CMD's for the
``high probability" thin disk (i.e. $TD/D<0.1$).
They are all essentially all like the CMD in Fig.~\ref{fig:cm_all}.

\paragraph{Thick disk ($TD/D>10$):}
The first thing that is evident from Fig.~\ref{fig:cm_many} is that the
proposed thick disk population have a prominent turn-off and that the CMD
essentially do not change when going from the lowest to the highest value
for the normalization. 
All stars that move into this $TD/D$ range when raising
the normalization come from the $1<TD/D<10$ bin.
Hence they are always classified as thick disk stars regardless
of the normalization.
The prominent turn-off that is seen is typical for an old stellar population
which is in concordance with the current beliefs of the Galactic thick disk
(see e.g. Fuhrmann~\cite{fuhrmann}).

\paragraph{Thick disk ($1<TD/D<10$):}
The CMD's for these stars resemble those for the stars with $TD/D>10$.
The number of stars in this $TD/D$ range almost double when going to 
the highest thick disk normalization. As in the case for the $TD/D>10$ stars
not much happens at first glance when raising the thick disk normalization.
However, when going from the 10\,\% to the 14\,\% normalization a few stars
with $M_{\rm V}$\,$<$\,5 and $B-V$\,$<$\,0.5 (i.e. in the upper left
area of the CMD) populate the CMD for the higher normalization.
These stars are most likely younger objects that should be attributed to the
thin disk. A 14\,\% normalization is therefore possibly too high 
given the velocity dispersions and rotational lags that we use
(see Table~\ref{tab:dispersions}).

\paragraph{Thin disk ($0.1<TD/D<1$):}
A young population such as the thin disk should include young stars,
especially the more massive stars that are still located on the main sequence.
For the 2\,\% normalization the CMD look suspiciously like the CMDs for the
samples with higher $TD/D$ ratios (i.e. the thick disk stars). The upper left
hand
area is poorly populated and the CMD probably mainly consists of stars with
``hot" thick disk kinematics that have a too low $TD/D$ ratio due to a too low
normalization. The picture is improved for the 6\,\% normalization
where a substantial number of young objects start to populate the CMD.\\

From this simple investigation we conclude that it is likely that a
normalization of 2\,\% is too low and a 14\,\% normalization probably
is too high for the thick disk in the solar neighbourhood.
Somewhere in between there is a dividing line where obviously
young objects starts to populate the thick disk CMD. The exact value for
this normalization is of course also dependent on the assumed velocity
dispersions in the disks. With our aaumptions it is however probably
located closer to 10\,\%. We have therefore
used $X_{\rm TD}$\,$=$\,10\,\% in the calculation of our
probabilities.

\paragraph{Selecting stars for abundance analysis:}
Where to put the limit where to assign a star as a ``true" thick disk
star is however difficult. The most safe way is to make it independent
of the thick disk normalization.
When doing this we will disregard the 2\,\% normalization since it is
obviously too low, and only consider values on $X_{\rm TD}$ between
6 and 14\,\%.
When calculating the $TD/D$ ratios with the 10\,\% normalization,
stars with $TD/D\gtrsim 2$ will then always have (independent on the
value on $X_{\rm TD}$) $TD/D>1$ and can be regarded as thick disk stars.
Stars with $TD/D\lesssim 0.6$ will always have (independent on the
value on $X_{\rm TD}$) $TD/D<1$ and can be regarded as thin disk stars.
Stars having $TD/D$ ratios in between these values will alter their
classification as thin or thick disk stars as the normalization increase or
decrease (in the range 6\,--\,14\,\%). We have therefore referred to 
these stars as ``transition objects" throughout the paper.

%==============================================================================

%==============================================================================
\end{document}